 \documentclass[final,3p,times,twocolumn]{elsarticle}

\usepackage{lipsum}
\usepackage{mathtools}
\usepackage{cuted}
\usepackage{amssymb}
\usepackage{amsmath}
\usepackage{lipsum}
\usepackage{graphicx}
\usepackage{float}
\usepackage[export]{adjustbox}
\usepackage{multicol}
\usepackage[normalem]{ulem}
\usepackage{xcolor}
\usepackage{hyperref}
\usepackage{tabularx}
\usepackage{breakurl}
\usepackage{hyperref} 
\usepackage{multibib}
\newcites{book}{Book References}
\newcites{journal}{Journal References}

\newcommand{\la}{\left\langle}
\newcommand{\ra}{\right\rangle}
\newcommand{\be}{\begin{equation}}
	\newcommand{\ee}{\end{equation}}
\newcommand{\bse}{\begin{subequations}}
	\newcommand{\ese}{\end{subequations}}
\newcommand{\bea}{\begin{eqnarray}}
	\newcommand{\eea}{\end{eqnarray}}
\newcommand{\ba}{\begin{array}}
	\newcommand{\ea}{\end{array}}

\journal{International Journal of Heat and Mass Transfer}

\begin{document}

\begin{frontmatter}

\title{Compressible turbulent convection at very high Rayleigh numbers}

\author[label1]{Harshit Tiwari\fnref{fn1}}
\address[label1]{Department of Physics, Indian Institute of Technology Kanpur, Kanpur 208016, India}


\author[label1]{Lekha Sharma\fnref{fn2}}

\author[label1]{Mahendra K. Verma\corref{cor1}
\fnref{fn3}}

\cortext[cor1]{Corresponding author}
\fntext[fn1]{tharshit@iitk.ac.in}
\fntext[fn2]{lekhasharma27@gmail.com}
\fntext[fn3]{mkv@iitk.ac.in}

\begin{abstract}
Heat transport in highly turbulent convection is not well understood. In this paper, we simulate compressible convection in a box of aspect ratio 4 using computationally-efficient MacCormack-TVD finite difference method on single and multi-GPUs, and   reach very high Rayleigh number (Ra)--- 10$\textsuperscript{15}$ in two dimensions and  10$\textsuperscript{11}$ in three dimensions.  We show that the Nusselt number Nu $\propto \mathrm{Ra}^{0.3}$ (classical scaling) that differs strongly from the ultimate-regime scaling, which is Nu $\propto \mathrm{Ra}^{1/2}$. The bulk temperature drops adiabatically along the vertical even for high Ra, which is in contrast to the constant bulk temperature in Rayleigh-B\'{e}nard convection (RBC). Unlike RBC, the density decreases with height. In addition, the vertical pressure-gradient ($-dp/dz$) nearly matches the buoyancy term ($\rho g$). But, the difference, $-dp/dz-\rho g$, is equal to the nonlinear term that leads to Reynolds number  $ \mathrm{Re} \propto  \mathrm{Ra}^{1/2}$. 

\end{abstract}


\end{frontmatter}



\section{Introduction}
\label{introduction}

Turbulent thermal convection occurs in a wide range of geophysical and astrophysical flows, as well as in many natural and industrial processes~\cite{Siggia:ARFM1994,Ahlers:RMP2009,Lohse:ARFM2010,Chilla:EPJE2012,Verma:book:BDF,Schumacher:RMP2020,Spruit:ARAA1990,Jones:Icarus2009}. There are various models of turbulent convection, the leading one being Rayleigh-B{\'e}nard convection (RBC)~\cite{Siggia:ARFM1994,Ahlers:RMP2009,Lohse:ARFM2010,Chilla:EPJE2012,Verma:book:BDF} that follows Oberbeck-Boussinesq (OB) approximation. In RBC, a horizontal layer of fluid confined between two parallel plates is heated from below and cooled from above. In the OB approximation, the density of the fluid is nearly constant, except for a small variation in the buoyancy term. For example, for water, whose thermal expansion coefficient $\alpha \approx 3\times 10^{-4}~\mathrm{K}^{-1}$ at room temperature, the relative change in density $(\delta \rho)/\rho \approx \alpha (\Delta T) \approx 10^{-3}$ when the temperature difference between the two plates is around 30 K~\cite{Verma:book:BDF}.

However, $(\delta \rho)/\rho$ in a turbulent star is of the order of unity \cite{Spiegel:AJ1965}. Let us assume that the star is made of ideal gas, whose thermal expansion coefficient is $1/T$, where $T$ is the gas temperature. Hence, $(\delta \rho)/\rho \approx \alpha (\Delta T) \approx 1$ when $(\Delta T) \approx T$. Therefore, the OB approximation and the RBC equations become invalid for stars and related astrophysical systems~\cite{Schumacher:RMP2020, Chandrasekhar:book:Instability, Spiegel:APJ1960, Hansen:book}. To study such systems, researchers often employ equations for compressible convection or intermediate models that employ non-Oberbeck-Boussinesq (NOB) effects~\cite{Wan:JFM2020,Pandey:PRF2021, Pandey:ApJ2021} or anelastic approximations~\cite{Gough:JAS1969, Verhoeven:AJ2015}. In this paper, we employ fully compressible equations to study turbulent convection at very large Rayleigh numbers (Ra)---up to $10^{15}$ for two-dimensions (2D), and $10^{11}$  for three-dimension (3D). Note, however, that the solar convection has even larger Ra ($\sim 10^{24}$) than the above~\cite{Schumacher:RMP2020}.

Many aspects of turbulent convection, which is more complex than hydrodynamic turbulence, are understood quite well, in particular for RBC systems. Grossmann and Lohse (GL in short) \cite{Grossmann:JFM2000, Grossmann:PRL2001} modelled the Reynolds number (Re) and Nusselt number (Nu) scaling using the relations for viscous and thermal dissipation rates \cite{Shraiman:PRA1990,Lesieur:book:Turbulence}. The GL relations have been verified by many experiments and numerical simulations~\cite{Stevens:JFM2013}. Recently, Bhattacharya and Verma~\cite{Bhattacharya:PoF2022} employed machine learning, artificial intelligence, and more accurate dissipation rates for Ra and Pr predictions. In addition, it has been shown that for small and moderate Prandtl numbers (Pr $\lessapprox 1$), turbulent convection has properties similar to hydrodynamic turbulence~\cite{Verma:book:BDF, Verma:NJP2017}. For example, the energy spectrum for turbulent convection follows Kolmogorov's 5/3 spectrum.

For the RBC setup, while the Reynolds number scaling  is reasonably well established, the properties  of velocity fluctuations remains to be quantified~\cite{Samuel:JFM2024}. Additionally, accurate characterization of the Nusselt number scaling remains a challenge. Two main theories describe the dependence of Nu on Ra. In the extreme turbulent regime,  known as  \textit{ultimate regime}, Kraichnan argued that Nu scales as $\mathrm{Ra}^{1/2}$~\cite{Kraichnan:PF1962Convection,Lohse:RMP2024}. On the contrary, Malkus~\cite{Malkus:PRSA1954} proposed that Nu is proportional to $\mathrm{Ra}^{1/3}$, referred to as the \textit{classical scaling}~\cite{Siggia:ARFM1994, Ahlers:RMP2009,  Lohse:ARFM2010, Chilla:EPJE2012, Grossmann:JFM2000,Lohse:RMP2024}. It is reported both in experiments and numerical simulations that up to $\mathrm{Ra}=10^{12}$, the Nu scaling exponent is near 0.30. However, Chavanne et al.~\cite{Chavanne:PF2001},  He et al.~\cite{He:PRL2012},  Zhu et al.~\cite{Zhu:PRL2018} and some others argued that the Nu exponent gradually increases to about 0.38 near $\mathrm{Ra} = 10^{15}$. Therefore, some researchers believe that the exponent might reach the value 1/2 at extreme Ra's~\cite{Lohse:RMP2024}. But, some others, based on different experiments and related simulations, argue that the classical scaling will remain valid for all Ra's~\cite{Schumacher:RMP2020, Niemela:Nature2000, Urban:PRL2012, Iyer:PNAS2020}. The scaling of Nu is discussed in detail in recent articles~\cite{Lohse:RMP2024, Sreenivasan:Atm2023}.

Turbulent compressible convection behaves very differently than RBC turbulence~\cite{John:PRF2023}. Unlike the constant bulk temperature in RBC, the bulk temperature in compressible convection falls adiabatically~\cite{Verhoeven:AJ2015}. In compressible convection, the fluid at the bottom is heavier than that at the top, which is in sharp contrast to that in RBC. Unfortunately, compressible convection has not been studied extensively. In recent times, Verhoeven et al.~\cite{Verhoeven:AJ2015}, and John and Schumacher~\cite{John:PRF2023,John:JFM2023,John:PF2024} analyzed turbulent compressible convection. Verhoeven et al.~\cite{Verhoeven:AJ2015} performed a comparative study of the anelastic approximation and fully compressible turbulent convection. John and Schumacher~\cite{John:PRF2023, John:JFM2023}
performed numerical simulations of fully compressible convection up to $\mathrm{Ra} = 10^7$ and explored different regimes of compressible convection. Some compressible simulations use inviscid flows~\cite{Porter:AJS2000} whose Ra estimates are somewhat uncertain. We find that there are no existing studies on fully compressible convection at very high Rayleigh numbers. The present paper aims to fill this gap.

In this paper, we simulate turbulent compressible convection using a  computationally-efficient MacCormack-TVD (total variation diminishing) finite difference method that eliminates numerical oscillations \cite{Ouyang:CG2013}. Our method is an extension of the novel and stable numerical scheme proposed by Ouyang et al. \cite{Ouyang:CG2013}, Yee \cite{yee:book}, and Liang et al. \cite{Liang:IJNMF2007} for shallow water equations to compressible turbulent convection. Our stable numerical scheme enabled us to perform convection simulations at very large Ra's, e.g., $10^{15}$ for 2D flows, and $10^{11}$ for 3D flows.

We quantify various quantities using the data generated by our numerical simulations. Our numerical data show that the internal energy dominates the fluid kinetic energy even at very large Ra. We attribute adiabaticity  to this effect.  In addition, we study the superadiabatic temperature and density of the flow. Based on numerical data, we report that $\mathrm{Nu} \sim \mathrm{Ra}^{0.3}$ up to $\mathrm{Ra} = 10^{15}$ in 2D and up to $10^{11}$ in 3D. We caution that our results differ significantly from those for RBC,  which is very different from compressible convection. Among the two, compressible equations model the astrophysical systems better.

Simulations of 3D turbulent convection at extreme Ra's are  very expensive. Fortunately, in the RBC framework, the scaling of Nu and Re  for 2D and 3D flows are reasonably similar for Pr ~$\gtrapprox 1$~\cite{Schmalzl:EPL2004, vanderPoel:JFM2013, Pandey:Pramana2016}. Motivated by these observations, we examine the Nu and Re scaling for extreme Ra in 2D, and for moderate Ra in 3D. We believe that these important findings will improve the models of stellar and planetary convection.

We have organised the paper as follows. In Sec.~\ref{sec:system}, we describe the physical system along with the governing equations. Sec.~\ref{sec:numerical_method} discusses the numerical scheme along with the simulation parameters. In Secs.~\ref{sec:flow_struc} and~\ref{sec:boundary_layer}, we describe the adiabaticity and boundary layers in turbulent convection. Section~\ref{sec:scaling} contains discussions on Nu and Re scaling. We conclude in Sec.~\ref{sec:summary}.\\

\begin{table*}[h]
\setlength{\tabcolsep}{2.8pt}
\centering
\begin{tabular}{|l p{0.5cm} r|}
\hline
\rule{0pt}{4ex} 
\textbf{Nomenclature} & & \\
 & & \\
\begin{tabular}{l l} 
$d$ & height of rectangular box \\
$L$ & length and width of rectangular box\\
$g$ & gravitational acceleration\\
$K$ & thermal conductivity\\
$T_b$ & temperature at the bottom plate\\
$T_t$ & temperature at the top plate\\
$C_p$ & specific heat capacity at constant pressure\\
$C_v$ & specific heat capacity at constant volume\\
$R_*$ & gas constant\\
$\tilde{\textbf{r}} = (\tilde{x},\tilde{y},\tilde{z})$ & dimensional position vector\\
$\textbf{r} = (x,y,z)$ & nondimensional position vector\\
$\tilde{\textbf{u}}$ & dimensional velocity\\
$\textbf{u}$ & nondimensional velocity\\
$\tilde{t}$ & dimensional time\\
$t$ & nondimensional time\\
$\tilde{T}$ & dimensional temperature\\
$T$ & nondimensional temperature\\
$\tilde{p}$ & dimensional pressure\\
$p$ & nondimensional pressure\\
$\tilde{E}$ & dimensional total energy density\\
$E$ & nondimensional total energy density\\
$\tilde{T}_A$ & dimensional adiabatic temperature\\
$T_A$ & nondimensional adiabatic temperature\\
$\tilde{p}_A$ & dimensional adiabatic pressure\\
$p_A$ & nondimensional adiabatic pressure\\
$\tilde{T}_{\mathrm{sa}}$ & dimensional superadiabatic temperature\\
$T_{\mathrm{sa}}$ & nondimensional superadiabatic temperature\\
$\mathrm{Pr}$ & Prandtl number\\
$\mathrm{Ra}$ & Rayleigh number\\
$\mathrm{Re}$ & Reynolds number\\
$D$ & dissipation number\\
$r = K_e/I_e$ & ratio of kinetic energy and internal energy\\
\end{tabular} &	 &
\begin{tabular}{l l}
    $\tilde{H}_T$ & dimensional total heat flux\\
    $\tilde{H}_A$ & dimensional adiabatic heat flux\\
    $\tilde{H}_{\mathrm{cond}}$ & dimensional conductive heat flux\\
    $\tilde{H}_{\mathrm{conv}}$ & dimensional convective heat flux\\
    $\overline{\mathrm{Nu}}$ & mean Nusselt number\\
    $\mathrm{Nu}$ & bulk Nusselt number\\
    $\mathrm{Nu}_{\mathrm{conv}}$ & convective Nusselt number\\
    $\mathrm{Nu}_K$ & $u_z$-induced kinetic Nusselt number\\
    $\tilde{U}$ & dimensional root-mean square velocity\\
    $U$ & nondimensional root-mean square velocity\\
    $\textbf{X}$ & column vector containing $\rho$, $\rho u_\alpha$, and $E$\\
    $\textbf{F}_\alpha$ & fluxes in $\alpha$-direction\\
    $\textbf{S}_\alpha$ & sources in $\alpha$-direction\\
    $\textbf{L}_\alpha$ & predictor-corrector operator $\alpha$-direction\\
    $\mathrm{Co}_\alpha$ & local Courant number in $\alpha$-direction\\ 
    $C(\mathrm{Ra})$ & correlation between $\rho u_z$ and $T_{\mathrm{sa}}$\\ \\
  \textit{Greek}: & \\
    $\Gamma$ & aspect ratio of computational domain\\
    $\Delta$ & temperature gradient in $z$-direction \\
    $\gamma$ & ratio of specific heats\\
    $\mu$ & dynamic viscosity\\
    $\nu$ & kinematic viscosity\\
    $\kappa$ & thermal diffusivity\\
    $\tilde{\rho}$ & dimensional fluid density\\
    $\rho$ & nondimensional fluid density\\
    $\tilde{\rho}_A$ & dimensional adiabatic density field\\
    $\rho_A$ & nondimensional adiabatic density field\\
    $\beta$ & adiabatic index\\
    $\epsilon$ & superadiabaticity\\
    $\tilde{\pmb{\tau}}$ & dimensional stress tensor\\
    $\pmb{\tau}$ & nondimensional stress tensor\\ \\
\end{tabular}\\
\hline
\end{tabular}
\end{table*}

\section{Physical System and Governing Equations}\label{sec:system}
We consider a fully compressible fluid confined in a rectangular box of dimension $(L, L, d)$ in 3D and $(L,d)$ in 2D, with the bottom and top plates at temperatures $T_b$  and $T_t$ respectively ($T_b > T_t$). Note that the adverse temperature gradient $\Delta = (T_b-T_t)/d>0$. We employ periodic boundary condition for the vertical sideways walls. However, for the top and bottom plates, we employ no-slip boundary conditions for the velocity field, and conducting boundary conditions  for the temperature field \cite{Verhoeven:AJ2015, John:JFM2023}. Note that a perfectly conducting plate maintains constant temperatures throughout its volume.

We assume that the fluid has constant dynamic viscosity $\mu$ and thermal conductivity $K$.  We also assume that the fluid follows  ideal gas law, $\tilde{p} = \tilde{\rho} R_* \tilde{T}$, where $\tilde{p}$ is the pressure; $\tilde{\rho}$ is the density of fluid; $R_* = C_p-C_v$ is the gas constant; $C_p, C_v$ are the specific heat capacities at constant pressure and volume respectively; and $\tilde{T}$ is the temperature field. The following conservative set of equations govern the system \cite{Spiegel:AJ1965, Graham:JFM1975}: 
{\begin{equation}
    \frac{\partial \tilde{\rho}}{\partial \tilde{t}} + \frac{\partial}{\partial \tilde{x}_i}(\tilde{\rho} \tilde{u}_i) = 0,\label{eq:continuity}
\end{equation}
\begin{equation}
    \frac{\partial}{\partial \tilde{t}}(\tilde{\rho} \tilde{u}_i) + \frac{\partial}{\partial \tilde{x}_j}(\tilde{\rho} \tilde{u}_i \tilde{u}_j + \delta_{ij}\tilde{p} - \tilde{\tau}_{ij}) = -g\tilde{\rho}\delta_{iz},\label{eq:momentum}
\end{equation}
\begin{equation}
    \frac{\partial \tilde{E}}{\partial \tilde{t}} + \frac{\partial}{\partial \tilde{x}_i} \left( \tilde{u}_i(\tilde{E} + \tilde{p}) - K \frac{\partial \tilde{T}}{\partial \tilde{x}_i} - \tilde{u}_j \tilde{\tau}_{ij} \right) = 0, \label{eq:energy}
\end{equation}}where $\tilde{u}_i$ are the velocity field components; $g$ is the acceleration due to gravity along $-\hat{z}$ direction;  
\be
\tilde{E} = \tilde{\rho} \left(\frac{\tilde{u}^2}{2} + C_v \tilde{T} + g\tilde{z} \right)
\ee
is the total energy density; and 
\be
\tilde{\tau}_{ij} = \mu \left( \tilde{\partial}_j \tilde{u}_i + \tilde{\partial}_i \tilde{u}_j + \frac{2}{3} \tilde{\partial}_m \tilde{u}_m \delta_{ij} \right)
\ee
is the stress tensor. Another important parameter is the aspect ratio $\Gamma$, which is the ratio of the length ($L$) and the height ($d$) of the rectangular box. All tilde variables have their respective dimensions.

The bulk flow is nearly adiabatic (or isentropic) because the convection time scale is faster than the conduction time scale (estimates in Sec.~\ref{sec:flow_struc}). Therefore, the vertical profiles of the adiabatic temperature $\tilde{T}_A(\tilde{z})$, adiabatic density $\tilde{\rho}_A(\tilde{z})$, and adiabatic pressure $\tilde{p}_A(\tilde{z})$ are~\cite{Spiegel:AJ1965,Verhoeven:AJ2015}
\begin{eqnarray}
    \tilde{T}_A(\tilde{z}) &=& \left(T_b - \frac{g}{C_p}\tilde{z} \right),
    \label{eq:T_adiab} \\
    \tilde{\rho}_A(\tilde{z}) &=& \frac{\rho_b}{T_b^\beta} (\tilde{T}_A(\tilde{z}))^\beta,\\
    \tilde{p}_A(\tilde{z}) &=& (C_p - C_v) \tilde{\rho}_A(\tilde{z}) \tilde{T}_A(\tilde{z}), 
\end{eqnarray}
where $T_b$ and $\rho_b$ are respectively the temperature and fluid density at the bottom plate; and $\beta =  1/(\gamma - 1)$ is the adiabatic index with $\gamma = C_p/C_v$. Later in the paper we will show the bulk temperature nearly follows the adiabatic profile [Eq.~(\ref{eq:T_adiab})]. The difference between the real temperature and the adiabatic profile is called \textit{superadiabatic temperature}: $\tilde{T}_\mathrm{sa}({\bf \tilde{r}},\tilde{t}) = \tilde{T}({\bf \tilde{r}}, \tilde{t})- \tilde{T}_A(\tilde{z})$.  

\begin{figure}[h]
    \centering
    \includegraphics[width = 0.48\textwidth]{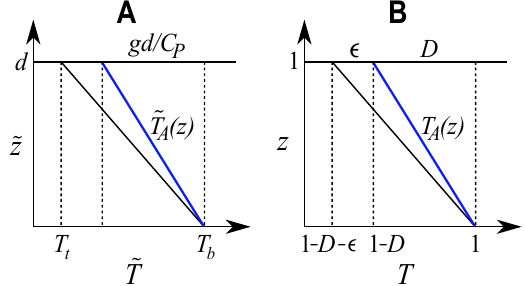}
    \caption{The adiabatic temperature gradient (blue lines) and the total temperature gradients (black lines) in (A) dimensional and (B) non-dimensional forms.
    }
    \label{fig:T_profile}
\end{figure}

Compressible convection has several important nondimensional parameters, which are~\cite{Verhoeven:AJ2015,John:PRF2023}
\setlength{\arraycolsep}{0.2em} 
\bea
\mathrm{superadiabaticity} ~\epsilon & = & \frac{d}{T_b} \left( \frac{\Delta}{d} - \frac{g}{C_p} \right), \\
\mathrm{dissipation~number} ~D & =& \frac{gd}{T_b C_p} = \frac{T_b-\tilde{T}_A(d)}{T_b}, \\
 \mathrm{Rayleigh~number} ~\mathrm{Ra} & = & \frac{\epsilon g d^3}{\nu \kappa}, 
 \label{eq:Ra_def} \\
 \mathrm{Prandtl~number} ~\mathrm{Pr} & =& \frac{\nu}{\kappa},\label{eq:Pr_def}
\eea
where $\nu = \mu/\rho$ is the kinematic viscosity, and $\kappa=K/(C_p\rho)$ is the thermal diffusivity of fluid. The parameters Ra and Pr are common with RBC, except that the temperature gradient has an adiabatic correction via $\epsilon$. The parameters $D$ and $\epsilon$ are unique to compressible convection: $D$ represents the nondimensional adiabatic temperature drop, whereas $\epsilon$ represents the excess temperature gradient relative to the adiabatic profile \cite{Verhoeven:AJ2015, John:PRF2023}. In Fig.~\ref{fig:T_profile}(A,B), we plot the total temperature gradient and the adiabatic temprature gradient in dimensional and nondimensional forms respectively.

In Sec.~\ref{sec:flow_struc}, we will show that the bulk temperature nearly follows the adiabatic temperature profile $\tilde{T}_A(\tilde{z})$, which is in sharp contrast to RBC where the bulk temperature is constant. Additionally, the density at the bottom is larger than that at the top, which is opposite to that in RBC. Therefore, compressible convection exhibits behavior that is markedly different from that of RBC.

We non-dimensionalize Eqs.~(\ref{eq:continuity})-(\ref{eq:energy}) using $d$ as the length scale, free-fall velocity $\sqrt{\epsilon g d}$ as the velocity scale, $\rho_b$ as the density scale, and $T_b$ as the temperature scale.  The governing equations in dimensionless form are \cite{Verhoeven:AJ2015,John:JFM2023}
\begin{equation}
     \frac{\partial \rho}{\partial t} + \frac{\partial}{\partial x_i}(\rho u_i) = 0,\label{eq:continuity_nondim}
\end{equation}
\begin{equation}
    \frac{\partial}{\partial t}(\rho u_i) + \frac{\partial}{\partial x_j}(\rho u_i u_j + \delta_{ij}p - \tau_{ij}) = -\frac{1}{\epsilon}\rho\delta_{iz},\label{eq:momentum_nondim}
\end{equation}
\begin{equation}
    \frac{\partial E}{\partial t} + \frac{\partial}{\partial x_i} \left( u_i(E + p) - \frac{1}{\epsilon D \sqrt{\mathrm{Ra} Pr}} \frac{\partial T}{\partial x_i} - u_j \tau_{ij} \right) = 0,\label{eq:energy_nondim}
\end{equation}
where the nondimensionalized pressure, total energy density, and stress tensor are defined respectively as:
\begin{eqnarray}
    p &=& \frac{(\gamma-1)}{\gamma \epsilon D} \rho T,\\
    E &=& \rho \left(\frac{u^2}{2} + \frac{1}{\gamma \epsilon D}T + \frac{1}{\epsilon}z \right), \\
    \tau_{ij} &=& \sqrt{\frac{\mathrm{Pr}}{\mathrm{Ra}}} \left( \partial_j u_i + \partial_i u_j + \frac{2}{3} \partial_m u_m \delta_{ij} \right).
\end{eqnarray}
Note that the tilde variables are dimensionful, whereas those without tilde are dimensionless. The nondimensional adiabatic profiles are \cite{Verhoeven:AJ2015}
\begin{eqnarray}
    T_A(z) &=& \left(1 - Dz \right),
    \label{eq:T_adiab_nd} \\
    \rho_A(z) &=& (1 - Dz)^\beta,\\
    p_A(z) &=& (1 - Dz)^{(\beta +1)}.
\end{eqnarray}
The nondimensional temperature is 1 at the bottom plate, and $1 - \epsilon - D$ at the top plate (see Fig.~\ref{fig:T_profile}). 

In this paper, we study heat transport  as a function of Ra for a fixed Pr, $\epsilon$, and $D$.  In the RBC framework, $\mathrm{Nu}$ is the ratio of total heat flux ($\tilde{H}_T$) and conductive heat flux ($\tilde{H}_{\mathrm{cond}}$). In compressible convection, the heat flux due to adiabatic cooling ($\tilde{H}_A$) must be subtracted from both $\tilde{H}_T$ and $\tilde{H}_{\mathrm{cond}}$ to compensate the thermodynamic effects~ \cite{Spiegel:APJ1960, Graham:JFM1975, Hurlburt:AJ1984}.  Here, the total heat flux is a sum of $\tilde{H}_{\mathrm{cond}} = K\Delta$ ($K$ is the thermal conductivity of the fluid), convective heat flux $\tilde{H}_{\mathrm{conv}} = C_p \langle \tilde{\rho} \tilde{u}_z \tilde{T}_{\mathrm{sa}} \rangle_{V,t}$, and $u_z$-induced kinetic heat flux $\tilde{H}_K = \langle \rho \tilde{u}_z \tilde{u}^2 / 2 \rangle_{V,t}$. The adiabatic heat flux is given by $\tilde{H}_A = Kg/C_p$. Hence, the Nusselt number is \cite{Graham:JFM1975, Hurlburt:AJ1984}
\begin{eqnarray}  
    \mathrm{Nu} &=& \frac{\tilde{H}_T - \tilde{H}_A}{\tilde{H}_{\mathrm{cond}} - \tilde{H}_A} \nonumber \\
    &=& \frac{K \Delta + C_p \langle \tilde{\rho} \tilde{u}_z \tilde{T}_{\mathrm{sa}} \rangle_{V,t} + \langle \rho \tilde{u}_z \tilde{u}^2 / 2 \rangle_{V,t} - Kg/C_p}{K \left(\Delta - \frac{g}{C_p} \right)} \nonumber \\
    &=& 1 + \frac{\sqrt{\mathrm{Ra}\mathrm{Pr}}}{\epsilon} \langle \rho u_z T_{\mathrm{sa}}\rangle_{V,t} + \frac{D \sqrt{\mathrm{Ra}\mathrm{Pr}}}{2} \langle \rho u_z u^2 \rangle_{V,t} \nonumber \\
    &=& 1 + \mathrm{Nu_{conv}} + \mathrm{Nu}_K. \label{eq:volume_Nu}
\end{eqnarray}
We report the volume and temporal averages of the above quantities and denote them by $\la . \ra_{V,t}$.

The above Nu, called \textit{bulk Nusselt number}, fluctuates significantly. In contrast, the Nusselt number computed near the bottom plate ($z=0$) and top plate ($z=1$) \cite{Verhoeven:AJ2015, John:JFM2023} 
\begin{equation}
    \mathrm{Nu}_{z=0,1} = -\frac{1}{\epsilon} \left. \frac{d \langle T_\mathrm{sa} \rangle_{A,t}}{dz} \right\vert_{z =0,1}, \label{eq:mean_Nu}
\end{equation}
have much less fluctuations. Here $\la . \ra_{A,t}$ represents the horizontal and temporal averages. Consequently, the mean Nusselt number at the boundaries is given by
\be
\overline{\mathrm{Nu}} = \frac{\mathrm{Nu}_{z=0} + \mathrm{Nu}_{z=1}}{2}.\label{eq:mean_Nu}
\ee
In this paper, we report Nu and $\overline{\mathrm{Nu}}$, as well as $\mathrm{Nu_{conv}}$ and $\mathrm{Nu}_K$. In addition to Nu, we also report the Reynolds number Re, which is defined as \cite{Verma:book:BDF}
\be
\mathrm{Re} = \frac{\tilde{U}d}{\nu}, \label{eq:Re_1}
\ee
where $\tilde{U}$ is the root-mean-square (rms) velocity. After non-dimensionalization \cite{Verhoeven:AJ2015},
\be
\mathrm{Re} = \sqrt{\frac{\mathrm{Ra}}{\mathrm{Pr}}} U = \sqrt{\frac{\mathrm{Ra}}{\mathrm{Pr}}} \langle \sqrt{\langle u^2\rangle_V}\rangle_t. \label{eq:Re}
\ee
where $U$ is the dimensionless velocity.

In Sec.~\ref{sec:numerical_method}, we solve Eqs.~(\ref{eq:continuity_nondim})-(\ref{eq:energy_nondim}) numerically.

\section{Numerical Method,  Validation, and Simulation Parameters}\label{sec:numerical_method}

In this section, we describe our numerical method, its validation, and the simulation parameters.

\subsection{Numerical Method} \label{subsec:method}
We solve Eqs.~(\ref{eq:continuity_nondim})-(\ref{eq:energy_nondim}) using  MacCormack-TVD (total variation diminishing) finite-difference scheme on a collocated grid \cite{Ouyang:CG2013, yee:book, Liang:IJNMF2007}. We employ a non-uniform tangent-hyperbolic grid in the $z$-direction to increase resolution near the boundaries, and uniform grids along the $x$ and $y$ directions.

All the equations are written in vectorial notation in the following conservative form \cite{Ouyang:CG2013}:
\begin{equation}
    \frac{\partial \textbf{X}}{\partial t} + \sum_\alpha \frac{\partial \textbf{F}_\alpha}{\partial x_\alpha} = \sum_\alpha \textbf{S}_\alpha,\label{eq:vector_eqn}
\end{equation}
where $\alpha = 1,2,3$ represent the $x$, $y$, and $z$ directions respectively. $\textbf{X}$ is a column vector that contains the variables $\rho$, $\rho u_\alpha$, and $E$; $\textbf{F}_\alpha$ and $\textbf{S}_\alpha$ are the respective  fluxes and sources \cite{Anderson:CFD1992}. Hence,
\begin{equation*}
    \textbf{X}=
    \begin{bmatrix}
    \rho\\
    \rho u_x\\
    \rho u_y\\
    \rho u_z\\
    E
    \end{bmatrix}
    ;
    \quad
    \textbf{S}_1=
    \begin{bmatrix}
    0\\
    f_1\\
    0\\
    0\\
    0
    \end{bmatrix}
    ;
    \quad
    \textbf{S}_2=
    \begin{bmatrix}
    0\\
    0\\
    f_2\\
    0\\
    0
    \end{bmatrix}
    ;
    \quad
    \textbf{S}_3=
    \begin{bmatrix}
    0\\
    0\\
    0\\
    f_3\\
    0
    \end{bmatrix}
    ;
\end{equation*}
\begin{equation}
    \textbf{F}_\alpha=
    \begin{bmatrix}
    \rho u_\alpha\\
    \rho u_x u_\alpha - \tau_{x \alpha}\\
    \rho u_y u_\alpha - \tau_{y \alpha}\\
    \rho u_z u_\alpha - \tau_{z \alpha} + p\\
    u_\alpha (E + p) - K \frac{\partial T}{\partial x_\alpha} - \sum_{\beta} u_\beta \tau_{\beta \alpha}
    \end{bmatrix}.
\end{equation}
For natural convection, $f_1 = 0$, $f_2 = 0$, and $f_3 = -\rho g$. Using operator-splitting method, we separate the Eq.~(\ref{eq:vector_eqn}) into three one-dimensional equations:
\begin{equation}
    \frac{\partial \textbf{X}}{\partial t} + \frac{\partial \textbf{F}_1}{\partial x_1} = \textbf{S}_1,
    \frac{\partial \textbf{X}}{\partial t} + \frac{\partial \textbf{F}_2}{\partial x_2} = \textbf{S}_2,
    \frac{\partial \textbf{X}}{\partial t} + \frac{\partial \textbf{F}_3}{\partial x_3} = \textbf{S}_3.
\end{equation}
The fields in $\textbf{X}$ are discretized in space, whose value at the site $(i,j,k)$ is $\textbf{X}_{ijk}$. We denote the time step using superscript $(n)$. The time stepping of $\textbf{X}_{ijk}$ from $(n)$ to $(n+1)$ is \cite{Ouyang:CG2013}
\begin{eqnarray}
    \textbf{X}_{ijk}^{(n+1)} = \textbf{L}_x \left( \frac{\delta t}{2} \right) \textbf{L}_y \left(\frac{\delta t}{2} \right) \textbf{L}_z \left(\frac{\delta t}{2} \right) \times \nonumber \\ \textbf{L}_x \left( \frac{\delta t}{2} \right) \textbf{L}_y \left(\frac{\delta t}{2} \right) \textbf{L}_z \left(\frac{\delta t}{2} \right) \textbf{X}_{ijk}^{(n)},\label{eq:time_step}
\end{eqnarray}
where $\delta t$ is the time-step; and $\textbf{L}_\alpha$ is predictor-corrector operator in $\alpha$-direction. The last three terms of Eq.~(\ref{eq:time_step}) carry forward $\textbf{X}$ from $(n)$ to $(n+1/2)$, whereas the first three terms take from $(n+1/2)$ to $(n+1)$, each of which involves several sub-steps.

Along the $z$-direction ($\alpha = 3$), a predictor step from $(n)$ to $(n+1/2)$ using backward difference is \cite{Ouyang:CG2013}
\begin{equation}
    \textbf{X}_{ijk}^P = \textbf{X}_{ijk}^{(n)} - \frac{\delta t}{2} \left( \frac{\textbf{F}_{\alpha,ijk}^{(n)} - \textbf{F}_{\alpha,ij(k-1)}^{(n)}}{\delta x_\alpha} \right) + \frac{\delta t}{2}\textbf{S}_{\alpha,ijk}^{(n)}.
\end{equation}
After this step, we perform the corrector step \cite{Ouyang:CG2013}
\begin{equation}
    \textbf{X}_{ijk}^C = \textbf{X}_{ijk}^{(n)} - \frac{\delta t}{2} \left( \frac{\textbf{F}_{\alpha,ij(k+1)}^{(n)} - \textbf{F}_{ijk}^{P}}{\delta x_\alpha} \right) + \frac{\delta t}{2}\textbf{S}_{\alpha,ijk}^{P}.
\end{equation}
We take an average of the above steps and add the TVD correction term $\textbf{T}_{ijk}^{(n+1/2)}$, which yields \cite{Ouyang:CG2013}
\begin{equation}
    \textbf{X}_{ijk}^{(n+1/2)} = \frac{\textbf{X}_{ijk}^{P} + \textbf{X}_{ijk}^{C}}{2} + \textbf{T}_{ijk}^{n+1/2},
\end{equation}
where
\begin{eqnarray}
    \textbf{T}_{ijk}^{n+1/2} &=& G(r_{ijk}^+ + r_{ij(k+1)}^-)~ \delta \textbf{X}_{ij(k+1/2)}^n - \nonumber \\ && G(r_{ij(k-1)}^+ + r_{ijk}^-)~ \delta \textbf{X}_{ij(k-1/2)}^n,
\end{eqnarray}
\begin{eqnarray}
    \delta \textbf{X}_{ij(k+1/2)}^n &=& \textbf{X}_{ij(k+1)}^n - \textbf{X}_{ijk}^n, \\ 
    \delta \textbf{X}_{ij(k-1/2)}^n &=& \textbf{X}_{ijk}^n - \textbf{X}_{ij(k-1)}^n,
\end{eqnarray}
\begin{eqnarray}
    r_{ijk}^\pm &=& \frac{\left(\delta \textbf{X}_{ij(k-1/2)}^n,\delta \textbf{X}_{ij(k+1/2)}^n \right)}{ \left(\delta \textbf{X}_{ij(k\pm1/2)}^n,\delta \textbf{X}_{ij(k\pm1/2)}^n \right)}.
\end{eqnarray}
The bracket $(\textbf{A},\textbf{B})$ indicates the dot product between $\textbf{A}$ and $\textbf{B}$, and
\begin{equation}
    G(x) = 0.5 C (1-\phi(x)),
\end{equation}
where $\phi(x) = \mathrm{max}(0,\mathrm{min}(2x,1))$ is the \textit{minmod flux limiter function}, and
\begin{equation}
    C = 
    \begin{cases}
        \mathrm{Co}_\alpha (1-\mathrm{Co}_\alpha), &\quad \mathrm{Co}_\alpha \le 0.5\\
        0.25, &\quad \mathrm{Co}_\alpha > 0.5
    \end{cases}
\end{equation}
with $\mathrm{Co}_\alpha$ representing the local Courant number in $\alpha$-direction \cite{Ouyang:CG2013, yee:book}. Identical processes are also adopted for the $x$- and $y$-directions.

The TVD correction preserves monotonicity and prevents spurious oscillations in the solution \cite{yee:book}. The MacCormack-TVD scheme is second-order accurate in space and time. For computing the boundary points, we use second-order forward and backward differences at the bottom and top plates, respectively.

\subsection{Programming Tools}

We developed an object-oriented Python solver, DHARA,  for simulating fully compressible equations on many GPUs and CPUs. The solver employs  CuPy library \cite{cupy_learningsys2017} for GPU acceleration, and mpi4py \cite{mpi4py} for multi-GPU communications. Note that  CuPy provides a GPU-optimized alternative to NumPy \cite{numpy}. We enhanced the code performance using \texttt{cupy.ElementwiseKernel()}, leading to 200X speedup on   NVIDIA A100 GPU in comparison to a single-core AMD EPYC 7543 processor. For scaling studies, we ran DHARA on  128 nodes (512 A100 GPUs) of Polaris  (Argonne National Laboratory)  and on 1024 nodes (8192 AMD MI250X GPUs) of Frontier (Oak Ridge National Laboratory) and demonstrated near ideal scaling (see~\ref{appendix1}). 

We performed the majority of our convection simulations on Polaris. Some simulations with smaller grids were performed on the Param Sanganak supercomputer of IIT Kanpur and on our laboratory clusters. 

\begin{table*}[h]
\setlength{\tabcolsep}{5.5pt} 
\caption{Comparison between our Nu and Re and those of John and Schumacher~\cite{John:JFM2023} for Ra = $10^6$, Pr = 0.7, $\Gamma = 4$, $\gamma = 1.4$, $\epsilon = 0.1$, and various $D$'s.}
\vspace{5pt}
\begin{tabular}{|c|c|c|c|c|c|c|c|}
\hline
Case & D & $\mathrm{Nu}$~\cite{John:JFM2023} & $\mathrm{Nu}$ (our simulations) & Error$_\mathrm{Nu}$ (in \%) & $\mathrm{Re}$~\cite{John:JFM2023} & $\mathrm{Re}$ (our simulations) & Error$_\mathrm{Re}$ (in \%) \\
\hline
1 & 0.1 & 7.94 & $7.90 \pm 0.06$ & 0.7 & 424 & $425 \pm 3$ & 0.7 \\
2 & 0.34 & 7.64 & $7.67 \pm 0.08$ & 1.0 & 414 & $414 \pm 4$ & 0.9 \\
3 & 0.5 & 6.93 & $6.89 \pm 0.05$ & 0.7 & 407 & $406 \pm 3$ & 0.7 \\
4 & 0.6 & 6.24 & $6.18 \pm 0.09$ & 1.4 & 370 & $368 \pm 5$ & 1.3 \\
\hline
\end{tabular}
\label{table:1}
\end{table*}
 
\begin{table*}[h]
\setlength{\tabcolsep}{5.5pt} 
\caption{Comparison between our Nu and Re and those of Verhoeven et al.~\cite{Verhoeven:AJ2015} for D = $0.49$, $\epsilon = 0.1$, Pr = $0.7$, $\Gamma = 2$, $\gamma = \frac{5}{3}$, and various Ra's.}
\vspace{5pt}
\begin{tabular}{|c|c|c|c|c|c|c|c|}
\hline
Case & $\mathrm{Ra}$ & $\mathrm{Nu}$~\cite{Verhoeven:AJ2015} & $\mathrm{Nu}$ (our simulations) & Error$_\mathrm{Nu}$ (in \%) & $\mathrm{Re}$~\cite{Verhoeven:AJ2015} & $\mathrm{Re}$ (our simulations) & Error$_\mathrm{Re}$ (in \%) \\
\hline
1 & $10^4$ & 2.1 & $2.175 \pm 0.005$  & 0.2 & 26.3 & $26.98 \pm 0.05$ & 0.2 \\
2 & $10^5$ & 3.9 & $4.0 \pm 0.1$ & 2.5 & 102 & $100 \pm 3$ & 3.3 \\
3 & $10^6$ & 7.1 & $7.3 \pm 0.2$ & 2.7 & 322 & $330 \pm 7$ & 2.0 \\
4 & $10^7$ & 13.4 & $14.1 \pm 0.4$ & 2.8 & 973 & $970 \pm 20$ & 2.4 \\
\hline
\end{tabular}
\label{table:2}
\end{table*}

\subsection{Validation}
We validate our code by comparing our numerical Nusselt and Reynolds numbers with those computed by John and Schumacher~\cite{John:JFM2023} and Verhoeven et al.~\cite{Verhoeven:AJ2015}. We compute Nu using Eq.~(\ref{eq:mean_Nu}). In particular, we simulate Eqs. (\ref{eq:continuity_nondim})-(\ref{eq:energy_nondim}) for Ra = $10^6$, $\epsilon = 0.1$, $\Gamma = 4$, $\gamma = 1.4$, Pr = $0.7$, and  $D=0.1, 0.34, 0.5, 0.6$. The Nu and Re computed by our code and those by John and Schumacher~\cite{John:JFM2023} are listed in Table~\ref{table:1}. Next, we compute Re and Nu by varying  $\mathrm{Ra}$ for a fixed  $D = 0.49$, $\Gamma = 2$, $\gamma = \frac{5}{3}$, $\epsilon = 0.1$,  $\mathrm{Pr} = 0.7$, and compare our results with Verhoeven et al.~\cite{Verhoeven:AJ2015} (see Table \ref{table:2}). Our results show good agreement with those reported by John and Schumacher~\cite{John:JFM2023} and Verhoeven et al.~\cite{Verhoeven:AJ2015}. Our Nu and Re differ from the previous works by less   than 3\%.

\begin{figure}[h]
    \centering
    \includegraphics[width = 0.48\textwidth]{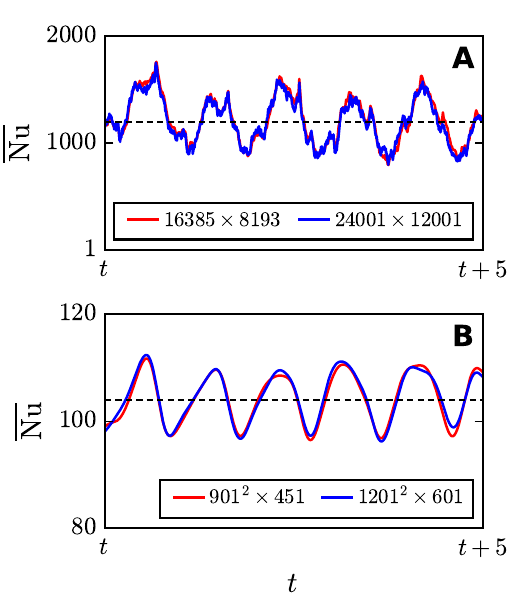}
    \caption{For $\epsilon=0.1, D=0.5, \gamma=1.3, \Gamma=4$, and $\mathrm{Pr}=0.7$, the time series of $\overline{\mathrm{Nu}}$ computed using Eq.~(\ref{eq:mean_Nu}) for (A) 2D convection at $\mathrm{Ra} = 10^{15}$ on grids $16385 \times 8193$ (red) and $24001 \times 12001$ (blue), and for (B) 3D convection at $\mathrm{Ra} = 10^{11}$ on grids $901^2 \times 451$ (red) and $1201^2 \times 601$ (blue).}
    \label{fig:nu_convergence}
\end{figure}
We also perform grid independence test by comparing Nu for 2D convection at $\mathrm{Ra}=10^{15}$ on grids $16385 \times 8193$ and $24001 \times 12001$, and for 3D convection at $\mathrm{Ra}=10^{11}$ on grids $901^2 \times 451$ and $1201^2 \times 601$. We compute $\overline{\mathrm{Nu}}$ for the steady states using Eq. (\ref{eq:mean_Nu}) and plot the time series in Fig.~\ref{fig:nu_convergence}(A, B) for 2D and 3D, respectively. The respective $\overline{\mathrm{Nu}}$ for 2D at $\mathrm{Ra}=10^{15}$ are 1218 and 1204, and for 3D at $\mathrm{Ra}=10^{11}$ are 103 and 104, which are within 1\% of each other. The near similarity of Nu time series and their averages indicate that our results are grid independent, as long as the flow is well resolved.

\subsection{Simulation Parameters}\label{sec:param}

We simulate Eqs.~(\ref{eq:continuity_nondim})-(\ref{eq:energy_nondim}) in 2D and 3D using the numerical scheme described in Sec. \ref{subsec:method}. For the horizontal plates, we employ no-slip boundary condition for the velocity field and conducting boundary condition for the temperature field. The fields at the side walls satisfy periodic boundary condition. 

For our simulations,  we fix $\Gamma = 4$, $\gamma = 1.3$, $\epsilon = 0.1$, $D = 0.5$, $\mathrm{Pr} = 0.7$ and vary Ra. The Rayleigh number Ra ranges from $10^9$ to $10^{15}$ for 2D, and from $10^8$ to $10^{11}$ for 3D. In Table \ref{table:3}, we list the grid sizes, Ra, $\overline{\mathrm{Nu}}$, Re, number of points in the top and bottom boundary layers ($N_t^{\mathrm{BL}}$ and $N_b^{\mathrm{BL}}$ respectively), and thicknesses of the top and bottom boundary layers ($\lambda_t$ and $\lambda_b$ respectively). 

We compute $\overline{\mathrm{Nu}}$ and Re  using  Eqs.~(\ref{eq:mean_Nu}, \ref{eq:Re}) respectively. However,  $\overline{\mathrm{Nu}}$ and Re  exhibit significant fluctuations, hence we average them over 50 time units in the steady state, except that we average over 30 and 20 time units for $\mathrm{Ra} = 10^{14}$ and $10^{15}$ respectively. We list these values in Table 3. Note that $\overline{\mathrm{Nu}}$ has around 10\% errors
for the 2D runs, but has maximum of 4\% error for the 3D
runs.  The errors in Re range from 1\% to 15\% for 2D flows, and from 0.5\% to 2\% for 3D flows.

The highest grid resolutions are $16385 \times 8193$ for 2D, and $901^2 \times 451$ for 3D. For all our runs, the number of grid points in the boundary layers exceeds 5, hence they satisfy the Gr{\"o}tzbach resolution criteria \cite{Lohse:RMP2024, Stevens:JFM2010, Grotzbach:JCP1983, Mishra:JFM2011}. These observations indicate that our simulations are well resolved.   For faster execution, we employ a single or mulitple GPUs: a single GPU for Runs 1-5 and 8; and 4 GPUs for Runs 6, 7 and 9-11  of Table \ref{table:3}. Our largest 3D run with $901^2 \times 451$ grid required $7$ days to complete on a Polaris node with four NVIDIA A100 GPUs.

\begin{table*}[h]
\centering
\caption{For our 2D convection Runs 1 to 7, and 3D Runs 8 to 11: the Rayleigh number $\mathrm{Ra}$, the grid size, the Reynolds number $\mathrm{Re}$, the mean Nusselt number  $\overline{\mathrm{Nu}}$, the number of grid points in the top and bottom boundary layers ($N_t^{\mathrm{BL}}$, $N_b^{\mathrm{BL}}$), and the thicknesses of top and bottom thermal boundary layers ($\lambda_t$, $\lambda_b$). Also, the polytropic index $\gamma=1.3$, Prandtl number $\mathrm{Pr} = 0.7$, dissipation number $D = 0.5$, superadiabaticity $\epsilon=0.1$, and the aspect ratio $\Gamma = 4$.}
\vspace{5pt}
\setlength{\tabcolsep}{9pt} 
\renewcommand{\arraystretch}{1.2} 
\begin{tabular}{|l|c|c|c|c|c|c|c|c|}
\hline
Run & Ra & Grid Size  & Re & $\overline{\mathrm{Nu}}$ & $N_t^{\mathrm{BL}}$ & $N_b^{\mathrm{BL}}$ & $\lambda_t$ & $\lambda_b$ \\
\hline
$1$ & $10^{9}$ & $3001 \times 1501$ & $(1.35 \pm 0.01) \times 10^{4}$ & $22 \pm 3$ & $133$ & $34$ & $0.055$ & $0.013$ \\
$2$ & $10^{10}$ & $5001 \times 2501$ & $(4.23 \pm 0.01) \times 10^{4}$ & $42 \pm 5$ & $180$ & $54$ & $0.036$ & $0.010$ \\
$3$ & $10^{11}$ & $5601 \times 2801$ & $(1.28 \pm 0.03) \times 10^{5}$ & $90 \pm 10$ & $120$ & $38$ & $0.020$ & $0.0062$ \\
$4$ & $10^{12}$ & $6001 \times 3001$ & $(4.04 \pm 0.07) \times 10^{5}$ & $180 \pm 20$ & $95$ & $32$ & $0.015$ & $0.0047$ \\
$5$ & $10^{13}$ & $8193 \times 4097$ & $(1.3 \pm 0.1) \times 10^{6}$ & $360 \pm 50$ & $70$ & $23$ & $0.010$ & $0.0031$ \\
$6$ & $10^{14}$ & $12001 \times 6001$ & $(4.0 \pm 0.2) \times 10^{6}$ & $570 \pm 70$ & $72$ & $21$ & $0.0066$ & $0.002$ \\
$7$ & $10^{15}$ & $16385 \times 8193$ & $(1.3 \pm 0.2) \times 10^{7}$ & $1200 \pm 200$ & $62$ & $18$ & $0.0042$ & $0.0012$ \\
\hline
$8$ & $10^{8}$ & $513^2 \times 257$ & $(2.21 \pm 0.01) \times 10^{3}$ & $14.0 \pm 0.2$ & $35$ & $12$ & $0.076$ & $0.0226$ \\
$9$ & $10^{9}$ & $801^2 \times 401$ & $(6.86 \pm 0.02) \times 10^{3}$ & $28.6 \pm 0.5$ & $32$ & $11$ & $0.047$ & $0.0158$ \\
$10$ & $10^{10}$ & $801^2 \times 401$ & $(2.03 \pm 0.04) \times 10^{4}$ & $53.9 \pm 0.5$ & $20$ & $7$ & $0.028$ & $0.010$ \\
$11$ & $10^{11}$ & $901^2 \times 451$ & $(5.86 \pm 0.08) \times 10^{4}$ & $103 \pm 4$ & $13$ & $5$ & $0.017$ & $0.0061$ \\
\hline
\end{tabular}
\label{table:3}
\end{table*}

In the following three sections, we will discuss various properties of compressible convection. We start with adiabatic profile and flow structures of the flow.

\section{Adiabatic Profile and Flow Structures} \label{sec:flow_struc}

According to \textit{Schwarzschild criterion},  at the onset of convection, the temperature in compressible convection  decreases vertically with the rate $g/C_p$, which is the adiabatic temperature profile~\cite{Landau:book:Fluid}. This result is derived using equilibrium thermodynamics. Interestingly, the atmospheres of the Earth and Sun too exhibit adiabatic temperature drop even though these systems are turbulent~\cite{Schumacher:RMP2020}. As we show below, we observe adiabatic cooling in our high Ra simulations, consistent with solar and Earth convection.
\begin{figure}[H]
    \centering
    \includegraphics[width = 0.48 \textwidth]{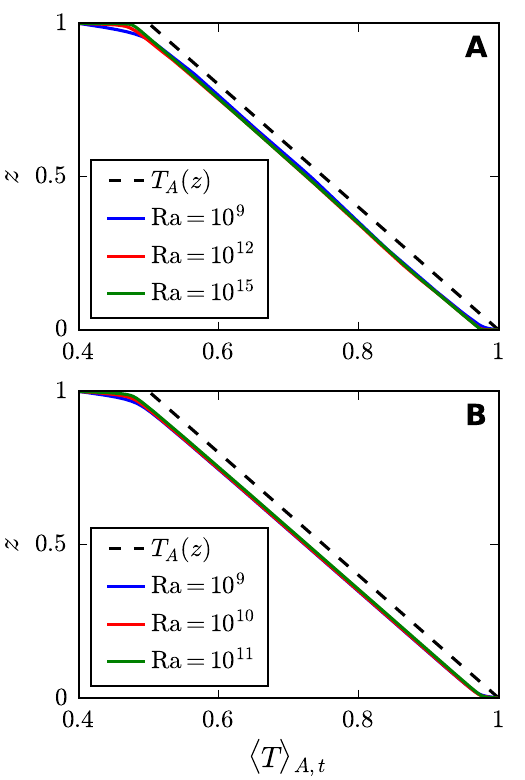}
    \caption{Profiles of horizontally and temporally averaged $\langle T \rangle_{A,t}(z)$ for (A) 2D convection for $\mathrm{Ra} = 10^9$ (blue), $\mathrm{Ra} = 10^{12}$ (red), and $\mathrm{Ra} = 10^{15}$ (green); and (B) 3D convection for $\mathrm{Ra} = 10^9$ (blue), $\mathrm{Ra} = 10^{10}$ (red), and $\mathrm{Ra} = 10^{11}$ (green). The black dashed line represents the adiabatic temperature ($T_A(z)$).}
    \label{fig:T_Ta_prof}
\end{figure}
\begin{figure}[H]
    \centering
    \includegraphics[width = 0.48 \textwidth]{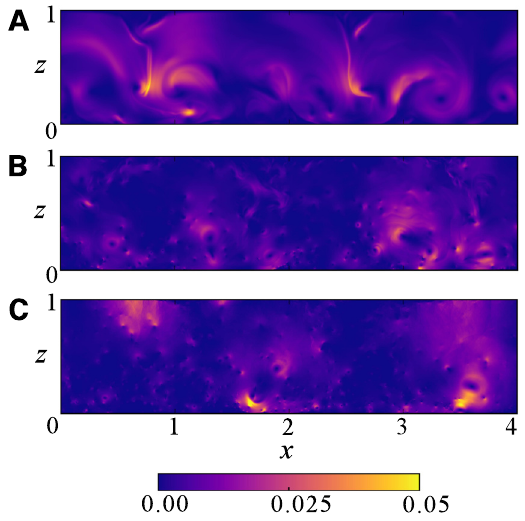}
    \caption{For 2D compressible convection with (A) $\mathrm{Ra} = 10^9$, (B) $10^{12}$, and (C) $10^{15}$, the density plots of the ratio of fluid kinetic energy and internal energy, $r =  (\rho u^2/2)/(\rho C_v T)$.}
    \label{fig:e_ratio}
\end{figure}
We compute horizontally and temporally averaged temperature $\la T \ra_{A,t}(z)$ for 2D and 3D runs. For a 2D flow, we average $T$ along the horizontal axis and over time during a steady state. In 3D, the corresponding averaging is performed over  horizontal planes and over time. In Fig.~\ref{fig:T_Ta_prof}(A, B), we plot $\la T \ra_{A,t}(z)$ for the 2D and 3D flows respectively. Interestingly, $dT(z)/dz = -g/C_p$ in the bulk, thus verifying adiabatic cooling. The reason for this observation is as follows.

\begin{figure}[h!]
    \centering
    \includegraphics[width = 0.48\textwidth]{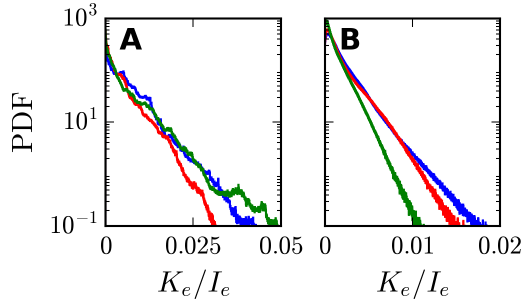}
    \caption{Normalized probability distribution function (PDF) of the ratio of the kinetic energy density, $K_e=\rho u^2/2$, and the internal energy density, $I_e = \rho C_v T$ for (A) 2D and (B) 3D convection during respective steady states. Color convention is the same as Fig.~\ref{fig:T_Ta_prof}.}
    \label{fig:pdf_e}
\end{figure}

We compute the ratio of the fluid kinetic energy and internal energy,  $r = K_e/I_e = (\rho u^2/2)/(\rho C_v T)$, at every grid point. For 2D simulation with $\mathrm{Ra} = 10^9, 10^{12}, 10^{15}$, Fig.~\ref{fig:e_ratio} illustrates the density plot of $r$. Figure~\ref{fig:pdf_e} illustrates the probability distribution function of $r$ for 2D and 3D convection. As shown in the figures, $r \to 0$, implying that the internal energy dominates the fluid kinetic energy, or $u \ll c_s$ ($c_s$ is the sound speed). The local Mach number of the flow is predominantly very small, but it reaches a maximum value of 0.9 at some isolated locations. The time average of maximum turbulent Mach number for all Ra's  vary from 0.5 to 0.6. This feature implies that the system is near thermodynamic equilibrium in the bulk. This is the reason why compressible convection exhibits adiabaticity in the bulk even for large Ra's \cite{Schumacher:RMP2020,Hansen:book}. These results are consistent with the fact that the convective time scale, $t_{\mathrm{conv}} = \sqrt{d/(\epsilon g)}$, is smaller than conductive time scale, $t_{\mathrm{cond}} = d^2/\kappa$. Their ratio $t_{\mathrm{conv}}/t_{\mathrm{cond}} = 1/\sqrt{\mathrm{Ra Pr}} \ll 1$. Note, however, that the Schwarzschild criterion does hold in the boundary layers, where the flow is dynamic.

\begin{figure}[h!]
    \centering
    \includegraphics[width = 0.48\textwidth]{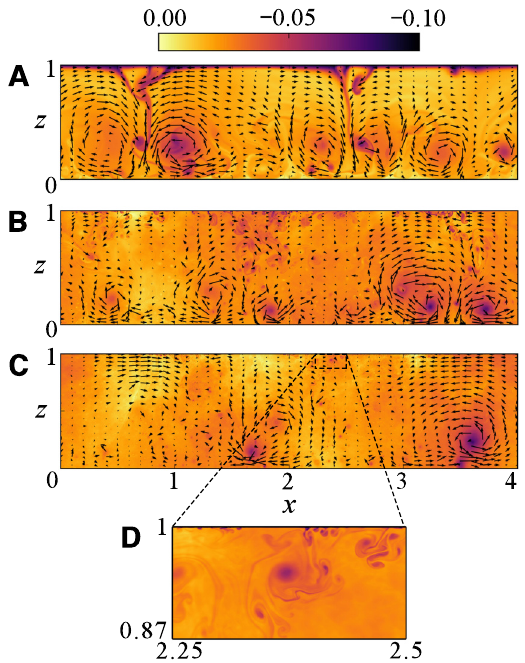}
    \caption{Plots of the velocity field $\bf{u}$ and superadiabatic temperature $T_{\mathrm{sa}}(\bf{r})$ for 2D convection with  (A) $\mathrm{Ra} = 10^9$, (B) $\mathrm{Ra} = 10^{12}$, and (C) $\mathrm{Ra} = 10^{15}$. (D) The magnified view near the top boundary for $\mathrm{Ra} = 10^{15}$.}
    \label{fig:field_profile_2d}
\end{figure}

\begin{figure}[H]
    \centering
    \includegraphics[width = 0.48\textwidth]{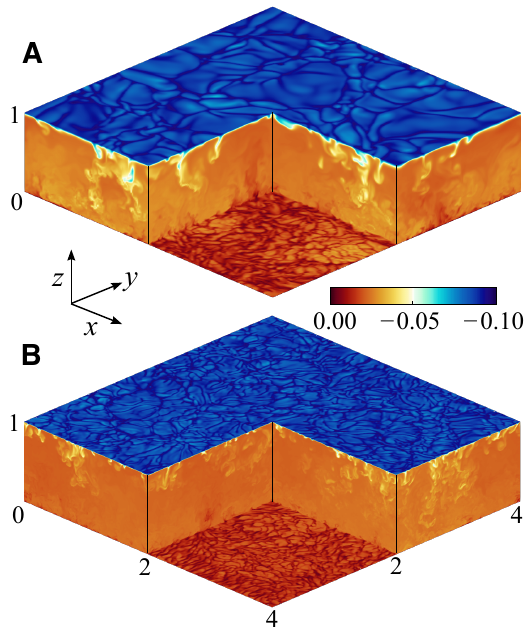}
    \caption{Plots of superadiabatic temperature $T_{\mathrm{sa}}(\bf{r})$ for 3D convection at (A) $\mathrm{Ra} = 10^9$ and (B) $\mathrm{Ra} = 10^{11}$.}
    \label{fig:field_profile_3d}
\end{figure}

Next, we discuss the flow structures of compressible convection. Since the temperature is dominated by $T_A(z)$, we subtract it from the total temperature and plot the superadiabatic temperature $T_{\mathrm{sa}}({\bf r}) = T({\bf r}) - T_A(z)$, along with the vector plots of the  velocity field. These plots are illustrated in Figs.~\ref{fig:field_profile_2d} and \ref{fig:field_profile_3d} for 2D and 3D flows respectively. For movies, refer to~\cite{movie}. The figures and movies exhibit flow activities at the bottom and the top of the box, where the thermal plumes are generated. Note, however, that the flow velocity of the plumes is much smaller than the sound speed.

As shown in Fig.~\ref{fig:T_Ta_prof}, the bulk temperature is nearly adiabatic, in contrast to  the constant bulk temperature in RBC. Interestingly, $T_\mathrm{sa} \ll T_A$ (see Figs.~\ref{fig:field_profile_2d} and \ref{fig:field_profile_3d}) indicating the dominance of $T_A$.  As shown in Figs.~\ref{fig:field_profile_2d} and \ref{fig:field_profile_3d}), turbulent convection is more prominent near the bottom than the top, consistent with earlier findings ~\cite{John:JFM2023,Wu:PRA1991}. As argued by John and Schumacher~\cite{John:JFM2023}, and Wu and Libchaber~\cite{Wu:PRA1991}, the fluid is denser at the bottom than the top. Hence, using $\mu = \nu\rho = $ const., we deduce that $\nu$ and $\kappa$ at the bottom are smaller than those at the top. Therefore,  the bottom region has stronger turbulence with relatively thin and easily-detachable plumes than the top region~\cite{John:PRF2023,Wu:PRA1991}. In the following, we discuss these  boundary layer features.

\section{Boundary Layers of Compressible Convection}\label{sec:boundary_layer}

\begin{figure}[H]
    \centering
    \includegraphics[width = 0.48\textwidth]{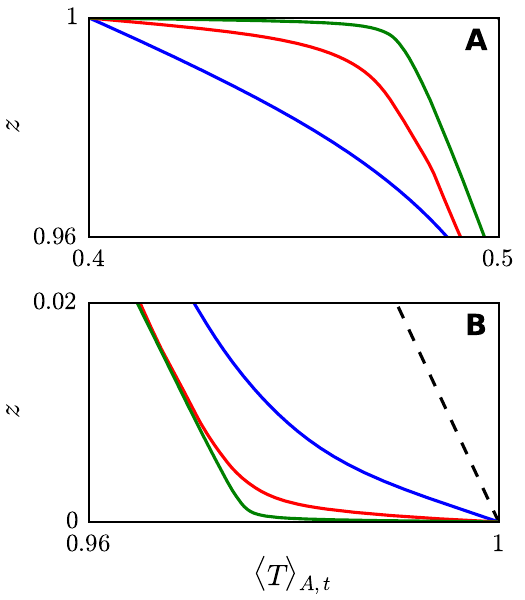}
    \caption{Boundary layers in  2D convection: The horizontally and temporally averaged $\langle T \rangle_{A,t}(z)$  in the (A) top boundary layer and (B) bottom boundary layer. Color convention is the same as Fig.~\ref{fig:T_Ta_prof}.}
    \label{fig:T_Ta_prof_zoom_2d}
\end{figure}

\begin{figure}[h]
    \centering
    \includegraphics[width = 0.48 \textwidth]{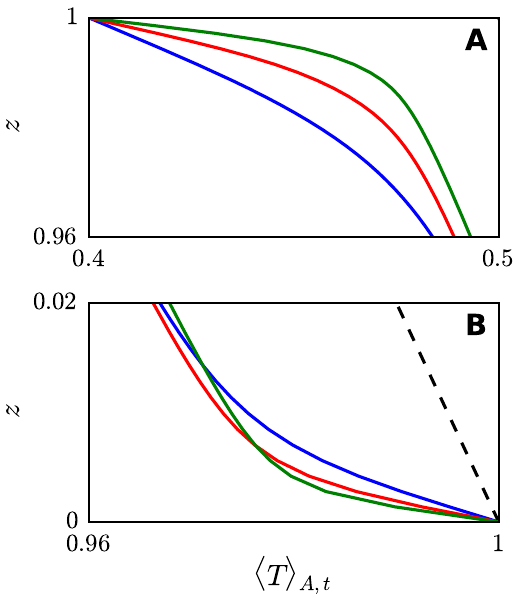}
    \caption{Boundary layers in  3D convection:  $\la T \ra_{A,t}(z)$  in the (A) top  boundary layer and (B) bottom boundary layer. Color convention is the same as Fig.~\ref{fig:T_Ta_prof}.}
    \label{fig:T_Ta_prof_zoom_3d}
\end{figure}

The planar-averaged temperature $\la T \ra_{A,t} (z) $, exhibited in Fig.~\ref{fig:T_Ta_prof}, shows that the temperature drops rather sharply at the top and bottom boundary layers. In Figs.~\ref{fig:T_Ta_prof_zoom_2d} and \ref{fig:T_Ta_prof_zoom_3d}, we plot the zoomed-view of the 2D and 3D boundary layers.

Unlike RBC, in compressible convection, the top boundary layer is thicker than the bottom one. At the bottom boundary layers of the 2D flows, the nondimensional temperature drops from 1 to $ 0.96$ (approximately) as $z$ increases from 0 to 0.02. However, at the top boundary layer, the temperature drops more rapidly---from 0.5 to 0.4 as $z$ varies from 0.96 to 1. A similar temperature drop is observed in 3D flows, albeit at low Ra's.  
 These abrupt temperature drops at the top are due to higher $\kappa$ at the top than the bottom because of the density profile~\cite{John:JFM2023,Wu:PRA1991}. Also note that at the top boundary layer, the rate of temperature drop  increases with Ra. The above results are consistent with those of John and Schumacher \cite{John:PRF2023, John:JFM2023}.

\begin{figure}[h]
    \centering
    \includegraphics[width = 0.48\textwidth]{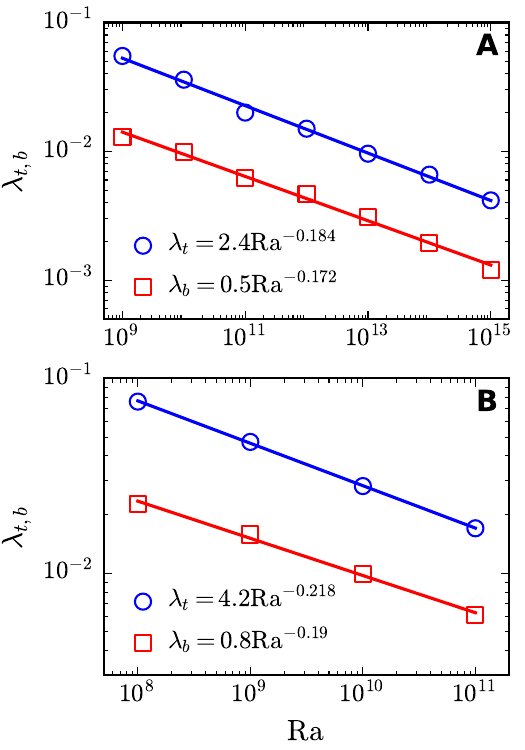}
    \caption{Plots of thermal boundary layer thicknesses $\lambda_{t,b}$ vs.~$\mathrm{Ra}$ for (A) 2D and (B) 3D flows. The blue and red symbols depict the thicknesses of the top and bottom boundary layers for various Ra's.}
    \label{fig:tbl}
\end{figure}

We compute the thicknesses of the top and bottom boundary layers. For the same, we identify  $z=\lambda_{t,b}$ where the slope of $\langle T \rangle_{A,t}$ approximates the adiabatic temperature gradient $-D$ near the top and bottom boundaries, respectively. The variation of $\lambda_{t,b}$ as a function of $\mathrm{Ra}$ is plotted in Fig.~\ref{fig:tbl}(A, B) for 2D and 3D flows. The red squares and blue circles represent the bottom and top boundary layers, respectively. It is evident that the top boundary layer is wider than the bottom boundary layer, and that the thickness of both boundary layers decrease with the increase of Ra. Interestingly, the boundary layer thickness follows a power law with Ra: $\lambda_t = (2.4 \pm 0.4) \mathrm{Ra}^{(-0.184 \pm 0.005)}, ~ \lambda_b = (0.5 \pm 0.1) \mathrm{Ra}^{(-0.172 \pm 0.006)}$ for 2D, and $\lambda_t = (4.2 \pm 0.2) \mathrm{Ra}^{(-0.218 \pm 0.003)}, ~ \lambda_b = (0.8 \pm 0.2) \mathrm{Ra}^{-0.19 \pm 0.01}$ for 3D. Note that the prefactor of $\lambda_t$ is larger than that of $\lambda_b$ because the top boundary layer is thicker than the bottom one. In contrast, the top and bottom boundary layers in  RBC scale identically, e.g., $\lambda_t \sim \lambda_b \sim \mathrm{Ra}^{-0.30}$ for $\mathrm{Pr} = 1$~\cite{Bhattacharya:PF2019}. Thus, boundary layers of RBC and compressible convection behave very differently.
 
\begin{figure}[h]
    \centering
    \includegraphics[width = 0.48\textwidth]{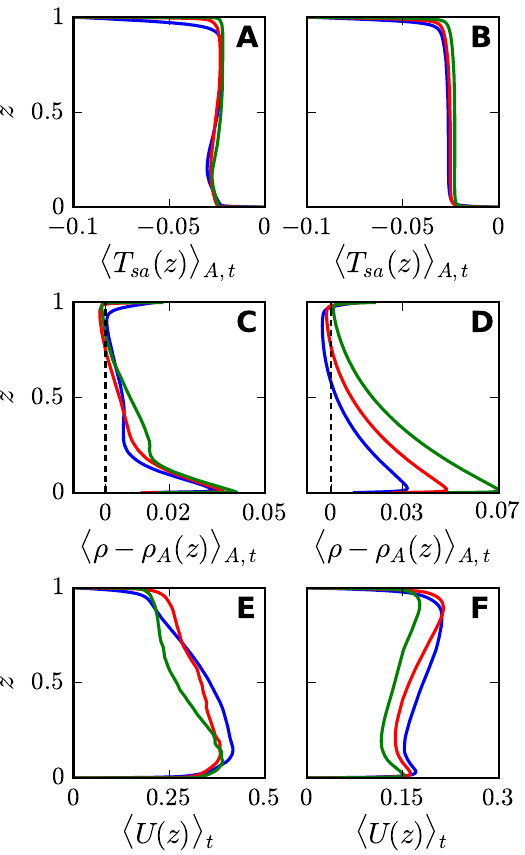}
    \caption{The planar and temporal averaged $T_{\mathrm{sa}}(z) = T(z) - T_A(z) $ (A,B), $\rho - \rho_A(z)$ (C,D), and rms velocity $U(z)$ (E,F). The left column represents 2D flows, whereas the right column represents  3D flows. 
    The black dashed lines in (C, D) denote adiabatic profile. Color convention is the same as   Fig.~\ref{fig:T_Ta_prof}.}
    \label{fig:vert_prof}
\end{figure}

In Fig.~\ref{fig:vert_prof} we plot the  temporal and planar averaged superadiabatic temperature $T_{\mathrm{sa}}(z) = T(z) - T_A(z)$, $\rho - \rho_A(z)$, and rms velocity $U(z)$. We observe that in the bulk, $T_{\mathrm{sa}}(z) \approx 0$ for both 2D and 3D because the bulk temperature nearly follows adiabatic profile (see Fig.~\ref{fig:vert_prof}(A, B)). The additional temperature gradients,  $dT_{\mathrm{sa}}(z)dz$, near the top and bottom boundary layers drive the flow. As shown in Fig.~\ref{fig:vert_prof}(A, B), $T_\mathrm{sa}(z)$ nearly collapses over each other, apart from some fluctuations.  As described earlier in this section, $T_\mathrm{sa}(z)$ drop is greater at the top than the bottom.

Figures~\ref{fig:vert_prof}(C, D) illustrate that the density is closer to the adiabatic profile in the upper box than in the lower box. The density variations from the adiabatic profile are greater near the bottom boundary than the top one. Note that the deviation from adibaticity for density ($\rho - \rho_A(z)$) increases  with Ra, both for 2D and 3D. Figures~\ref{fig:vert_prof}(E, F) illustrate $\la U(z) \ra_t$ arising due to the turbulent activities. 
Note that $\la U(z) \ra_t$ decreases with increase in $\mathrm{Ra}$ for both 2D and 3D flows. Also, for 2D convection, $\la U(z) \ra_t$ at the bottom is larger than that near the top. However, it is nearly reversed in 3D.  Thus, $\la U(z) \ra_t$  exhibits asymmetry  around the mid-plane, which is unlike nearly symmetric $\la U(z) \ra_t$ in RBC~\cite{Mishra:PRE2010}.


In the next section, we will discuss Nu and Re scaling for turbulent compressible convection.

\section{Nu and Re Scaling for Compressible Convection}
\label{sec:scaling}

In this section, we compute the global measures, Re and Nu,  for 2D and 3D compressible convection.

\subsection{\rm{Re} Scaling}\label{subsec:Re}
\begin{figure}[h]
    \centering
    \includegraphics[width = 0.48\textwidth]{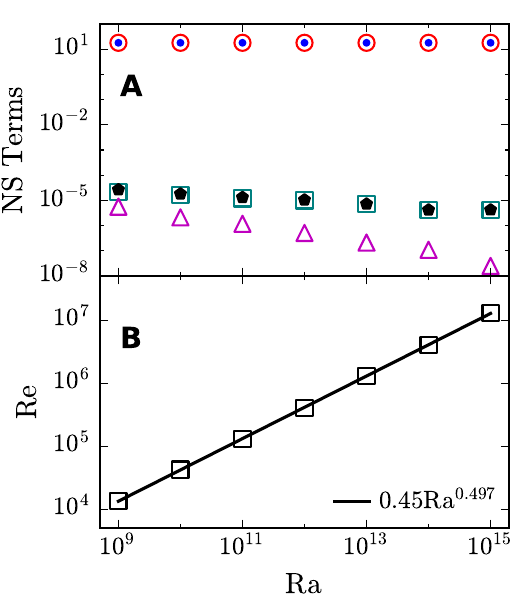}
    \caption{For 2D convection, along the $z$ direction of the momentum equation: (A) the volume averaged $\langle \mathrm{F}_p \rangle$ (red unfilled circles), $\langle \mathrm{F}_b \rangle$ (blue dots), $\langle \mathrm{F}_n \rangle$ (teal unfilled squares), $\langle \mathrm{F}_v \rangle$ (magenta unfilled triangles), and $\langle \mathrm{F}_b \rangle - \langle \mathrm{F}_p \rangle$ (black pentagons); (B) the Re scaling, which is Re $\sim \mathrm{Ra}^{0.497}$.}
    \label{fig:reynolds_2d}
\end{figure}
\begin{figure}[h]
    \centering
    \includegraphics[width = 0.48\textwidth]{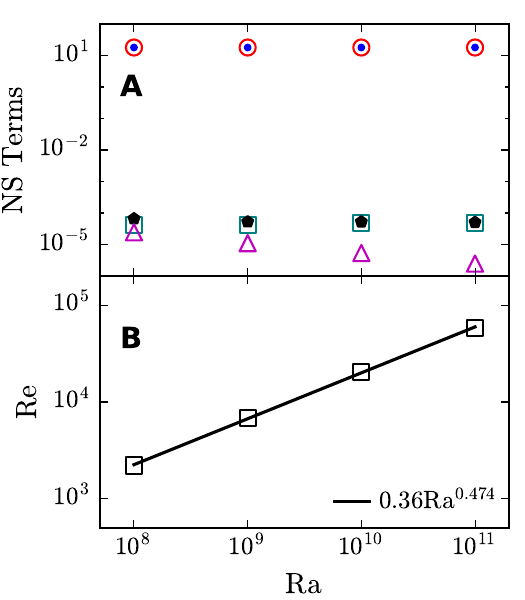}
    \caption{For 3D convection, along the $z$ direction of the momentum equation: (A) the volume averaged $\langle \mathrm{F}_p \rangle$ (red unfilled circles), $\langle \mathrm{F}_b \rangle$ (blue dots), $\langle \mathrm{F}_n \rangle$ (teal unfilled squares), $\langle \mathrm{F}_v \rangle$ (magenta unfilled triangles), and $\langle \mathrm{F}_b \rangle - \langle \mathrm{F}_p \rangle$ (black pentagons); (B)  the Re scaling, which is Re $\sim \mathrm{Ra}^{0.474}$.}
    \label{fig:reynolds_3d}
\end{figure}

We analyze the relative strengths of various terms of the momentum equation along $\hat{z}$~[Eq.~(\ref{eq:momentum})]. The different terms of the equation along $\hat{z}$ are
\begin{equation}
    \mathrm{Nonlinear~term}: \mathrm{F}_n = \rho (\mathbf{u} \cdot \pmb{\nabla}) u_z,
\end{equation}
\begin{equation}
    \mathrm{Pressure~gradient}: \mathrm{F}_p = -\frac{\partial p}{\partial z},
\end{equation}
\begin{equation}
    \mathrm{Buoyancy}: \mathrm{F}_b = \rho g,
\end{equation}
\begin{equation}
    \mathrm{Viscous~force}: \mathrm{F}_v = \partial_i\tau_{iz}.
\end{equation}
We compute the volume averages of the above four terms, along with  $\langle \mathrm{F}_b \rangle - \langle \mathrm{F}_p \rangle $. These quantities are plotted in Figs.~(\ref{fig:reynolds_2d}A, \ref{fig:reynolds_3d}A) for 2D and 3D flows respectively. The plots clearly show that 
\be
\langle \mathrm{F}_p \rangle \approx \langle \mathrm{F}_b \rangle.
\label{eq:pressure_grav_balance}
\ee
This balance is a key feature of \textit{Schwarzschild criterion}~\cite{Landau:book:Fluid} (see Sec.~\ref{sec:flow_struc}). Interestingly, Eq.~(\ref{eq:pressure_grav_balance}) is satisfied for larger Ra's as well. This is the reason why bulk flow of compressible convection is nearly adiabatic for high Ra's as well.

It turns out that $\langle \mathrm{F}_p \rangle$ and $\langle \mathrm{F}_b \rangle$ do not cancel precisely, and the difference between the two terms nearly equals  the  nonlinear term, i.e.,
\be
\langle \mathrm{F}_n \rangle  \approx \langle \mathrm{F}_b \rangle - \langle \mathrm{F}_p \rangle.
\label{eq:F_n_balance}
\ee
The viscous term, which is  $\lessapprox 10^{-5}$ for all Ra's, is much less than the other three terms.  Using Eq.~(\ref{eq:F_n_balance}) we derive that 
\begin{equation}
    \frac{\tilde{\rho} \tilde{U}^2}{d} \sim \epsilon \tilde{\rho} g,
\end{equation}
where $\epsilon$ is superadiabaticity, a small constant. Using Eq.~(\ref{eq:Ra_def}) and~(\ref{eq:Pr_def}), we derive that
\begin{equation}
    \mathrm{Re} = \frac{\tilde{U} d}{\nu} \approx \sqrt{\frac{\mathrm{Ra}}{\mathrm{Pr}}}.\label{eq:reynolds}
\end{equation}
Therefore, $\mathrm{Re} \propto \mathrm{Ra}^{1/2}$.   We remark that the momentum equation of RBC exhibits a different balance among various terms, with significant weight for the nonlinear term~\cite{Verma:book:BDF, Pandey:PRE2016}. Consequently, the Re scaling for the two systems are somewhat different. The relation $\mathrm{Re} \propto \mathrm{Ra}^{1/2}$ remains the dominant scaling for RBC as well~\cite{Grossmann:JFM2000}, but the exponent correction may reach 0.05 for large Ra's ~\cite{Verma:book:BDF, Grossmann:JFM2000}.

Using the 2D and 3D numerical data, we compute $\mathrm{Re}$ for various Ra's in the respective steady states, and list them in Table~\ref{table:3} (see Sec.~\ref{sec:param}). The listed Re's have errors ranging from 1\% to 15\% for the 2D flows, and from 0.5\% to 2\% for the 3D flows. We exhibit the Re vs.~Ra plots for the 2D and 3D flows in Figs.~(\ref{fig:reynolds_2d}B, \ref{fig:reynolds_3d}B). We observe that $\mathrm{Re} = (0.45 \pm 0.02) \mathrm{Ra}^{(0.497 \pm 0.002)}$ for 2D, and $\mathrm{Re} = (0.36 \pm 0.04) \mathrm{Ra}^{(0.474 \pm 0.005)}$ for 3D.  These numerical results are consistent with the  prediction that $\mathrm{Re} \propto \mathrm{Ra}^{1/2}$.
  
\subsection{\rm{Nu} Scaling}\label{subsec:Nu}
Here, we discuss the Nu scaling. We compute the volume- and time-averaged $\mathrm{Nu}$, convection-induced $\mathrm{Nu_{conv}}$, and $u_z$-induced $\mathrm{Nu}_K$ using Eq.~(\ref{eq:volume_Nu}), as well as boundary-layer $\overline{\mathrm{Nu}}$ using Eq.~(\ref{eq:mean_Nu}) for 2D and 3D runs. We employ a moving average over 10 free-fall times. The bulk Nu's, the first three quantities of the above, are listed in Table~\ref{table:4}, whereas $\overline{\mathrm{Nu}}$ is listed in Table~\ref{table:3}.  Note that $\overline{\mathrm{Nu}}$, averaged over nearly 50 time units, has a maximum of 17\% error for the 2D runs, but a maximum of 4\% errors for the 3D runs (see Table~\ref{table:3} and Sec.~\ref{sec:param}).

\begin{table}[H]
\centering
\caption{The volume- and time-averaged (over 10 free-fall times) $\mathrm{Nu}$, $\mathrm{Nu}_\mathrm{conv}$ and   $\mathrm{Nu}_K$ of Eq.~(\ref{eq:volume_Nu}) for 2D and 3D runs. These quantities have significant fluctuations, especially in 2D convection. $\mathrm{Nu}_K$ is relatively small with large fluctuations. In the table, we do not report $\mathrm{Nu}_K$ for runs 2 to 5 because of large fluctuations. }
\vspace{5pt}
\setlength{\tabcolsep}{4.8pt} 
\begin{tabular}{|l|c|c|c|c|}
\hline
Run & Ra & Nu & Nu\textsubscript{conv} & Nu\textsubscript{\textit{K}} \\
\hline
$1$ & $10^{9}$ & $21 \pm 2$ & $28 \pm 2$ & $-6 \pm 2$ \\
$2$ & $10^{10}$ & $39 \pm 7$ & $52 \pm 4$ & - \\
$3$ & $10^{11}$ & $80 \pm 23$ & $114 \pm 8$ & - \\
$4$ & $10^{12}$ & $160 \pm 50$ & $250 \pm 30$ & - \\
$5$ & $10^{13}$ & $300 \pm 150$ & $500 \pm 100$ & - \\
\hline
$6$ & $10^{8}$ & $13.9 \pm 0.4$ & $16.1 \pm 0.5$ & $-1.9 \pm 0.7$ \\
$9$ & $10^{9}$ & $27 \pm 2$ & $32 \pm 2$ & $-4 \pm 1$ \\
$7$ & $10^{10}$ & $55 \pm 2$ & $64 \pm 3$ & $-9 \pm 1$ \\
$8$ & $10^{11}$ & $110 \pm 14$ & $130 \pm 15$ & $-18 \pm 2$ \\
\hline
\end{tabular}
\label{table:4}
\end{table}

\begin{figure}[h]
    \centering
    \includegraphics[width = 0.48\textwidth]{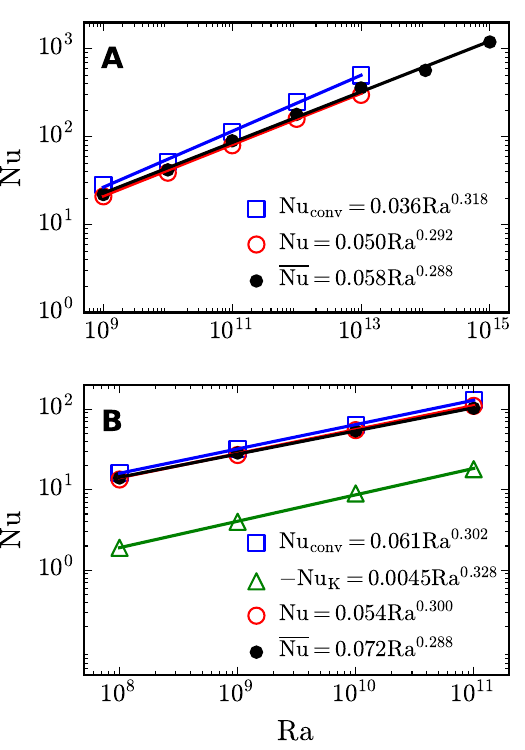}
    \caption{Nu Scaling for (A) 2D and (B) 3D convection. The figure exhibits $\mathrm{Nu}_\mathrm{conv}$, $\mathrm{Nu}_K$, $\mathrm{Nu}$, and $\overline{\mathrm{Nu}}$ [see Eqs.~(\ref{eq:volume_Nu}, \ref{eq:mean_Nu})]. Note that Nu follows near classical scaling in both 2D and 3D.}
    \label{fig:nusselt}
\end{figure}

The bulk Nu fluctuates significantly for 2D convection~\cite{vanderPoel:PRE2014}, hence we time average them over approximately 10 time units. The total Nu's have large errors, especially for large Ra's. We exclude the bulk Nu's for 2D at $\mathrm{Ra}=10^{14}$ and $\mathrm{Ra}=10^{15}$ as they exhibit more than 100\% error. The errors in Nu for the 2D flows range from 10\% up to 50\%, but they are within 10\% for the 3D flows. We observe that $\mathrm{Nu}_\mathrm{conv}$ has similar errors as Nu, but  $\mathrm{Nu}_K$ has relatively large errors for 2D convection. We do not list $\mathrm{Nu}_K$ for Runs 2 to 5 due to large errors.
Fortunately, $\mathrm{Nu}_K \ll \mathrm{Nu}_\mathrm{conv}$, hence we can ignore  $\mathrm{Nu}_K$ safely.

In Figs.~\ref{fig:nusselt}(A, B), we plot the averaged Nu, $\mathrm{Nu}_\mathrm{conv}$, and $\overline{\mathrm{Nu}}$ for 2D and 3D flows. We also plot $\mathrm{Nu}_K$ for 3D flows. Interestingly, $\overline{\mathrm{Nu}} \approx \mathrm{Nu}$, but Nu has much larger fluctuations than  $\overline{\mathrm{Nu}}$. Also, $\mathrm{Nu}_K \ll  \mathrm{Nu_{conv}}$ up to $\mathrm{Ra} = 10^{13}$. Hence, in Eq.~(\ref{eq:volume_Nu}), the maximal contribution to  $\mathrm{Nu}$ comes from $\mathrm{Nu_{conv}}$. The fluid is denser at the bottom than near the top, which leads to $\rho u_z u^2$ being more negative than positive. Hence, $\mathrm{Nu}_K \sim \langle \rho u_z u^2 \rangle <0$ in compressible convection. This is unlike RBC where $u_z$ is symmetric around the mid-plane, leading to $\langle \rho u_z u^2 \rangle = 0$~\cite{Siggia:ARFM1994, Ahlers:RMP2009}. 

Quantitatively, $\mathrm{Nu_{conv}} = (0.036 \pm 0.007) \mathrm{Ra}^{(0.318 \pm 0.007)}$ for 2D convection, and $ (0.061 \pm 0.002) \mathrm{Ra}^{(0.302 \pm 0.002)}$  for 3D convection. The volume-averaged Nu  scales as $(0.050 \pm 0.005) \mathrm{Ra}^{(0.292 \pm 0.004)}$ for 2D, and $(0.054 \pm 0.003) \mathrm{Ra}^{(0.300 \pm 0.003)}$ for 3D. We observe that $\overline{\mathrm{Nu}}$ follows a similar scaling as Nu. Thus, turbulent convection follows \textit{classical scaling}, rather than \textit{ultimate-regime scaling}.

\begin{figure}[H]
    \centering
    \includegraphics[width = 0.48\textwidth]{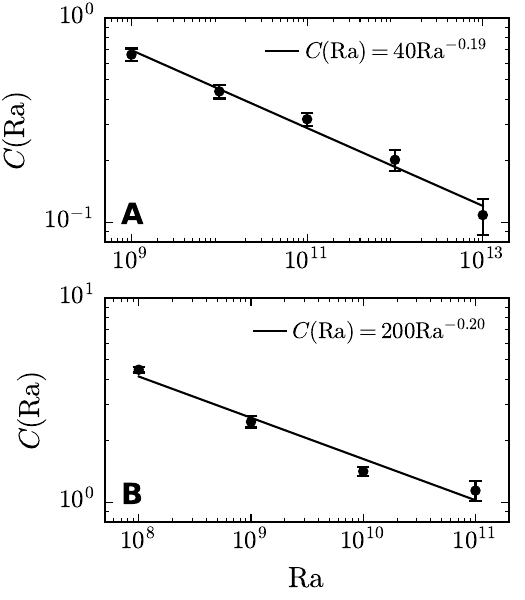}
    \caption{Plots of correlation $C(\mathrm{Ra})$ [Eq.~(\ref{eq:Corr_Ra})]  for (A) 2D and (B) 3D convection. These correlations bring down the Nu scaling from 1/2 to near 0.3.}
    \label{fig:correlations}
\end{figure}

The above results lead to [see Eq.~(\ref{eq:volume_Nu})]
\be
\mathrm{Nu} \approx \mathrm{Nu}_\mathrm{conv} \approx \la \tilde{\rho} \tilde{u}_z \tilde{T}_\mathrm{sa} \ra \approx 
C(\mathrm{Ra}) \sqrt{\la (\tilde{\rho} \tilde{u}_z)^2 \ra \la \tilde{T}_\mathrm{sa}^2 \ra} ,
\label{eq:Nu_corr}
\ee
where 
\be
C(\mathrm{Ra}) = \frac{\la \tilde{\rho} \tilde{u}_z \tilde{T}_\mathrm{sa} \ra }{\sqrt{\la (\tilde{\rho} \tilde{u}_z)^2 \ra \la \tilde{T}_\mathrm{sa}^2 \ra}}
\label{eq:Corr_Ra}
\ee
is the normalized correlation between $\tilde{\rho} \tilde{u}_z$ and $\tilde{T}_\mathrm{sa}$.
Note that $\sqrt{\la (\tilde{\rho} \tilde{u}_z)^2 \ra \la \tilde{T}_\mathrm{sa}^2 \ra} \sim U \sim \sqrt{\mathrm{Ra Pr}}$.  The correlation $C(\mathrm{Ra})$ plotted in  Fig.~\ref{fig:correlations} reveals that $C(\mathrm{Ra}) = (40 \pm 10) \mathrm{Ra}^{(-0.19 \pm 0.01)}$ for 2D, and $C(\mathrm{Ra}) = (200 \pm 100) \mathrm{Ra}^{(-0.20 \pm 0.03)}$ for 3D. Hence, $C(\mathrm{Ra})$ corrects the Nu exponent from 1/2 to near 0.3.  This feature is similar to those observed by Verma et al. \cite{Verma:PRE2012} for RBC. However, the component $\mathrm{Nu}_K$ present in compressible convection, which is negative, further suppresses the Nu exponent to $0.29$ in 2D and $0.3$ in 3D.

Even though the dynamics of RBC and compressible convection are significantly different, the classical Re and Nu scaling are very similar. This feature is possibly due to similar $C$(Ra) scaling in both the systems, an issue that needs a closer examination. 

\section{Summary}\label{sec:summary}


In this paper, we present simulation results of compressible convection for very high Ra's. We simulated 2D and 3D convective flows for $\mathrm{Pr} = 0.7$, superadiabaticity parameter $\epsilon = 0.1$,  and dissipation number $D = 0.5$. We choose Ra's in the range of $10^9$ to $10^{15}$ in 2D, but $\mathrm{Ra} = 10^8$, $10^9$, $10^{10}$ and $10^{11}$ in 3D. The main results reported in the paper are as follows:
\begin{enumerate}
    \item  For all our runs, the pressure gradient nearly matches with the buoyancy, thus satisfying the adiabaticity condition or Schwarzschild criterion, even for very large Ra. The density too is nearly adiabatic with the fluid density decreasing with height.  We show that the adiabaticity arises because the internal energy is much stronger than the fluid kinetic energy. Hence, the flow is in quasi thermodynamic equilibrium, except near the bottom and top plates where the temperature gradients exceed $g/C_P$. 

    \item Unlike RBC, the flow and the boundary layers of compressible convection are asymmetric along the vertical.  For example, the top boundary layer is thicker than the bottom boundary. With the increase in $\mathrm{Ra}$, the thicknesses of both the thermal boundary layers decrease with Ra as $\sim \mathrm{Ra}^{-0.178}$ in 2D and as $\sim \mathrm{Ra}^{-0.2}$ in 3D. 

    \item For high Ra's, the pressure gradient and buoyancy do not cancel each other exactly.  The difference between the two terms nearly equals the nonlinear term. Using this feature, we derive that $\mathrm{Re} \sim \mathrm{Ra}^{1/2}$. Our numerical results nearly follow the above scaling.

    \item We show that the Nusselt number for compressible convection exhibits near classical scaling ($\mathrm{Ra}^{0.30}$) up to Ra = $10^{15}$ in 2D and up to $10^{11}$ in 3D. Note that for RBC, several researchers~\cite{He:PRL2012,Lohse:RMP2024} have reported a gradual transition to the ultimate regime, which appears to be absent in compressible convection, at least up to Ra $=10^{15}$ for 2D convection. 

    \item Many features of compressible convection deviate  from RBC, where the temperature and the density in the bulk are nearly constant. Despite these differences, the Re and Nu scaling for compressible convection and RBC are quite similar. 

\end{enumerate}

Our results on compressible convection are of major importance to the  atmospheres of the Earth and the Sun, both of which exhibit near adiabatic temperature profile while being turbulent. We plan to perform somewhat realistic simulations of the Earth and the Sun in near future; the latter simulation would require higher Rayleigh numbers ($10^{22}$ to $10^{24}$) and inclusion of the magnetic field. These studies would provide us valuable insights into the energetics, heat transport, and magnetic field dynamics in the Sun.

\section*{Acknowledgements}
The authors thank K. R. Sreenivasan and J{\"o}rg Schumacher for useful discussions. We also thank Argonne Leadership Computing Facility (ALCF) and Oak Ridge National Laboratory (ORNL) for the computer time through the Director’s Discretionary Program.  Simulations were performed on Polaris, Frontier,  Param Sanganak, and our laboratory GPUs. LS acknowledges the Institute Postdoctoral Fellowship of IITK. Part of this work was supported by the Science and Engineering Research Board, India (Grant Nos. SERB/PHY/2021522 and SERB/PHY/2021473), and the J. C. Bose Fellowship (SERB /PHY/2023488).

\appendix

\section{Scalability of DHARA}
\label{appendix1}
\begin{figure}[h]
    \centering
    \includegraphics[width = 0.48\textwidth]{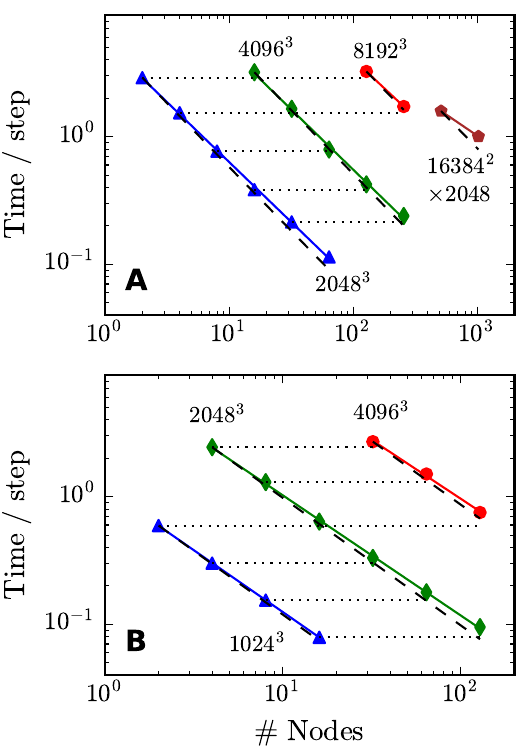}
    \caption{Scalability of DHARA on GPUs of (A) Frontier and (B) Polaris. The figure shows the time taken per timestep vs. the number of nodes, demonstrating strong and weak scaling.}
    \label{fig:scaling_dhara}
\end{figure}
We performed scaling analysis of DHARA (finite-difference solver) on  Frontier of Oak Ridge National Laboratory (OLCF) and on Polaris of Argonne Leadership Computing Facility (ALCF). Each node of Frontier contains four AMD MI250X, each with 2 Graphics Compute Dies (GCDs), whereas each node of Polaris contains four NVIDIA A100 GPUs. On Frontier, we vary grids from $2048^3$ to $16384^2 \times 2048$ and nodes  from 8 to 1024. On Polaris, the corresponding grids were varied  from $1024^3$ to $4096^3$, and the nodes  from 2 to 128.

Figure \ref{fig:scaling_dhara}(A, B) illustrate the time taken for a timestep as a function of number of nodes on Frontier and Polaris respectively. The reported time is averaged over several timesteps. We observe that the time taken $T \propto n^{-1}$, where $n$ is the number of nodes, thus indicating strong scaling for DHARA in both systems. In addition, DHARA shows good weak scaling because the time taken remains unchanged when the grid size and number of nodes are increased proportionally. 



\bibliographystyle{ieeetr}


\begin{thebibliography}{10}
	
	\bibitem{Siggia:ARFM1994}
	E.~D. Siggia, ``{High Rayleigh number convection},'' {\em Annu. Rev. Fluid
		Mech.}, vol.~26, no.~1, pp.~137--168, 1994.
	
	\bibitem{Ahlers:RMP2009}
	G.~Ahlers, S.~Grossmann, and D.~Lohse, ``{Heat transfer and large scale
		dynamics in turbulent Rayleigh-B{\'e}nard convection},'' {\em Rev. Mod.
		Phys.}, vol.~81, pp.~503--537, Apr. 2009.
	
	\bibitem{Lohse:ARFM2010}
	D.~Lohse and K.-Q. Xia, ``{Small-scale properties of turbulent
		Rayleigh{\textendash}B{\'e}nard convection},'' {\em Annu. Rev. Fluid Mech.},
	vol.~42, no.~1, pp.~335--364, 2010.
	
	\bibitem{Chilla:EPJE2012}
	F.~Chill{\`a} and J.~Schumacher, ``{New perspectives in turbulent
		Rayleigh-B{\'e}nard convection},'' {\em Eur. Phys. J. E}, vol.~35, no.~7,
	p.~58, 2012.
	
	\bibitem{Verma:book:BDF}
	M.~K. Verma, {\em Physics of Buoyant Flows: From Instabilities to Turbulence}.
	\newblock Singapore: World Scientific, 2018.
	
	\bibitem{Schumacher:RMP2020}
	J.~Schumacher and K.~R. Sreenivasan, ``Colloquium: Unusual dynamics of
	convection in the {S}un,'' {\em Rev. Mod. Phys.}, vol.~92, no.~4, p.~041001,
	2020.
	
	\bibitem{Spruit:ARAA1990}
	H.~C. Spruit, A.~Nordlund, and A.~M. Title, ``{Solar convection},'' {\em Annu.
		Rev. Astron. Astrophys.}, vol.~28, no.~1, pp.~263--303, 1990.
	
	\bibitem{Jones:Icarus2009}
	C.~A. Jones and K.~M. Kuzanyan, ``Compressible convection in the deep
	atmospheres of giant planets,'' {\em Icarus}, vol.~204, no.~1, pp.~227--238,
	2009.
	
	\bibitem{Spiegel:AJ1965}
	E.~A. Spiegel, ``{Convective instability in a compressible atmosphere. I.},''
	{\em Astrophys. J.}, vol.~141, pp.~1068--1090, 1965.
	
	\bibitem{Chandrasekhar:book:Instability}
	S.~Chandrasekhar, {\em {Hydrodynamic and Hydromagnetic Stability}}.
	\newblock Clarendon: Oxford University Press, 1961.
	
	\bibitem{Spiegel:APJ1960}
	E.~A. Spiegel and G.~Veronis, ``On the {B}oussinesq approximation for a
	compressible fluid.,'' {\em Astrophys. J.}, vol.~131, p.~442, 1960.
	
	\bibitem{Hansen:book}
	C.~J. Hansen, S.~D. Kawaler, and V.~Trimble, {\em {Stellar Interiors: Physical
			Principles, Structure, and Evolution}}.
	\newblock New York: Springer-Verlag, 2012.
	
	\bibitem{Wan:JFM2020}
	Z.-H. Wan, Q.~Wang, B.~Wang, S.~Xia, Q.~Zhou, and D.-J. Sun, ``On
	non-{O}berbeck--{B}oussinesq effects in {R}ayleigh--{B}{\'e}nard convection
	of air for large temperature differences,'' {\em J. Fluid Mech.}, vol.~889,
	p.~A10, 2020.
	
	\bibitem{Pandey:PRF2021}
	A.~Pandey, J.~Schumacher, and K.~R. Sreenivasan, ``Non-{B}oussinesq convection
	at low {P}randtl numbers relevant to the {S}un,'' {\em Phys. Rev. Fluids},
	vol.~6, no.~10, p.~100503, 2021.
	
	\bibitem{Pandey:ApJ2021}
	A.~Pandey, J.~Schumacher, and K.~R. Sreenivasan, ``{Non-Boussinesq
		low-Prandtl-number convection with a temperature-dependent thermal
		diffusivity},'' {\em Astrophys. J.}, vol.~907, no.~1, p.~56, 2021.
	
	\bibitem{Gough:JAS1969}
	D.~O. Gough, ``{The anelastic approximation for thermal convection},'' {\em J.
		Atmos. Sci.}, vol.~26, pp.~448--456, May 1969.
	
	\bibitem{Verhoeven:AJ2015}
	J.~Verhoeven, T.~Wieseh{\"o}fer, and S.~Stellmach, ``Anelastic versus fully
	compressible turbulent {R}ayleigh--{B}{\'e}nard convection,'' {\em Astrophys.
		J.}, vol.~805, no.~1, p.~62, 2015.
	
	\bibitem{Grossmann:JFM2000}
	S.~Grossmann and D.~Lohse, ``{Scaling in thermal convection: a unifying
		theory},'' {\em J. Fluid Mech.}, vol.~407, pp.~27--56, Mar. 2000.
	
	\bibitem{Grossmann:PRL2001}
	S.~Grossmann and D.~Lohse, ``{Thermal convection for large Prandtl numbers},''
	{\em Phys. Rev. Lett.}, vol.~86, pp.~3316--3319, Apr. 2001.
	
	\bibitem{Shraiman:PRA1990}
	B.~I. Shraiman and E.~D. Siggia, ``{Heat transport in high-Rayleigh-number
		convection},'' {\em Phys. Rev. A}, vol.~42, no.~6, pp.~3650--3653, 1990.
	
	\bibitem{Lesieur:book:Turbulence}
	M.~Lesieur, {\em {Turbulence in Fluids}}.
	\newblock Dordrecht: Springer-Verlag, 2008.
	
	\bibitem{Stevens:JFM2013}
	R.~J. A.~M. Stevens, E.~P. van~der Poel, S.~Grossmann, and D.~Lohse, ``{The
		unifying theory of scaling in thermal convection: the updated prefactors},''
	{\em J. Fluid Mech.}, vol.~730, pp.~295--308, July 2013.
	
	\bibitem{Bhattacharya:PoF2022}
	S.~Bhattacharya, M.~K. Verma, and A.~Bhattacharya, ``{Predictions of Reynolds
		and Nusselt numbers in turbulent convection using machine-learning models},''
	{\em Phys. Fluids}, vol.~34, no.~2, 2022.
	
	\bibitem{Verma:NJP2017}
	M.~K. Verma, A.~Kumar, and A.~Pandey, ``{Phenomenology of buoyancy-driven
		turbulence: recent results},'' {\em New J. Phys.}, vol.~19, p.~025012, 2017.
	
	\bibitem{Samuel:JFM2024}
	R.~J. Samuel, M.~Bode, J.~D. Scheel, K.~R. Sreenivasan, and J.~Schumacher, ``No
	sustained mean velocity in the boundary region of plane thermal convection,''
	{\em Journal of Fluid Mechanics}, vol.~996, p.~A49, 2024.
	
	\bibitem{Kraichnan:PF1962Convection}
	R.~H. Kraichnan, ``{Turbulent thermal convection at arbitrary Prandtl
		number},'' {\em Phys. Fluids}, vol.~5, pp.~1374--1389, Nov. 1962.
	
	\bibitem{Lohse:RMP2024}
	D.~Lohse and O.~Shishkina, ``{Ultimate Rayleigh-B{\'e}nard turbulence},'' {\em
		Rev. Mod. Phys.}, vol.~96, no.~3, p.~035001, 2024.
	
	\bibitem{Malkus:PRSA1954}
	W.~V.~R. Malkus, ``{The Heat Transport and Spectrum of Thermal Turbulence},''
	{\em Proc. R. Soc. Lond. A}, vol.~225, pp.~196--212, Aug. 1954.
	
	\bibitem{Chavanne:PF2001}
	X.~Chavanne, F.~Chill{\`a}, B.~Chabaud, B.~Castaing, and B.~Hebral,
	``{Turbulent Rayleigh-B{\'e}nard convection in gaseous and liquid He},'' {\em
		Phys. Fluids}, vol.~13, pp.~1300--1320, May 2001.
	
	\bibitem{He:PRL2012}
	X.~He, D.~Funfschilling, H.~Nobach, E.~Bodenschatz, and G.~Ahlers,
	``{Transition to the Ultimate State of Turbulent Rayleigh-B{\'e}nard
		Convection},'' {\em Phys. Rev. Lett.}, vol.~108, p.~024502, Jan. 2012.
	
	\bibitem{Zhu:PRL2018}
	X.~Zhu, V.~Mathai, R.~J. A.~M. Stevens, R.~Verzicco, and D.~Lohse,
	``{Transition to the Ultimate Regime in Two-Dimensional Rayleigh-B{\'e}nard
		Convection},'' {\em Phys. Rev. Lett.}, vol.~120, p.~144502, Apr. 2018.
	
	\bibitem{Niemela:Nature2000}
	J.~J. Niemela, L.~Skrbek, K.~R. Sreenivasan, and R.~J. Donnelly, ``{Turbulent
		convection at very high Rayleigh numbers},'' {\em Nature}, vol.~404,
	pp.~837--840, Jan. 2000.
	
	\bibitem{Urban:PRL2012}
	P.~Urban, P.~Hanzelka, T.~Kralik, V.~Musilov{\'a}, A.~Srnka, and L.~Skrbek,
	``{Effect of Boundary Layers Asymmetry on Heat Transfer Efficiency in
		Turbulent Rayleigh-B{\'e}nard Convection at Very High Rayleigh Numbers},''
	{\em Phys. Rev. Lett.}, vol.~109, p.~154301, Oct. 2012.
	
	\bibitem{Iyer:PNAS2020}
	K.~P. Iyer, J.~D. Scheel, J.~Schumacher, and K.~R. Sreenivasan, ``{Classical
		1/3 scaling of convection holds up to Ra$~= 10^{15}$},'' {\em Proc. Natl.
		Acad. Sci.}, vol.~117, no.~14, pp.~7594--7598, 2020.
	
	\bibitem{Sreenivasan:Atm2023}
	K.~R. Sreenivasan and J.~J. Niemela, ``{Turbulent convection at very high
		Rayleigh numbers and the weakly nonlinear theory},'' {\em Atmosphere},
	vol.~14, no.~5, p.~826, 2023.
	
	\bibitem{John:PRF2023}
	J.~P. John and J.~Schumacher, ``Strongly superadiabatic and stratified limits
	of compressible convection,'' {\em Phys. Rev. Fluids}, vol.~8, no.~10,
	p.~103505, 2023.
	
	\bibitem{John:JFM2023}
	J.~P. John and J.~Schumacher, ``Compressible turbulent convection in highly
	stratified adiabatic background,'' {\em J. Fluid Mech.}, vol.~972, p.~R4,
	2023.
	
	\bibitem{John:PF2024}
	J.~P. John and J.~Schumacher, ``{Compressible turbulent convection: The role of
		temperature-dependent thermal conductivity and dynamic viscosity},'' {\em
		Phys. Fluids}, vol.~36, no.~7, 2024.
	
	\bibitem{Porter:AJS2000}
	D.~H. Porter and P.~R. Woodward, ``Three-dimensional simulations of turbulent
	compressible convection,'' {\em Astrophys. J. Suppl. Ser.}, vol.~127, no.~1,
	p.~159, 2000.
	
	\bibitem{Ouyang:CG2013}
	C.~Ouyang, S.~He, Q.~Xu, Y.~Luo, and W.~Zhang, ``{A MacCormack-TVD finite
		difference method to simulate the mass flow in mountainous terrain with
		variable computational domain},'' {\em Comput. Geosci.}, vol.~52, pp.~1--10,
	2013.
	
	\bibitem{yee:book}
	H.~C. Yee, {\em Upwind and symmetric shock-capturing schemes}.
	\newblock National Aeronautics and Space Administration, Ames Research Center,
	1987.
	
	\bibitem{Liang:IJNMF2007}
	D.~Liang, B.~Lin, and R.~A. Falconer, ``{Simulation of rapidly varying flow
		using an efficient TVD-MacCormack scheme},'' {\em Int. J. Numer. Methods
		Fluids}, vol.~53, no.~5, pp.~811--826, 2007.
	
	\bibitem{Schmalzl:EPL2004}
	J.~Schmalzl, M.~Breuer, and U.~Hansen, ``{On the validity of two-dimensional
		numerical approaches to time-dependent thermal convection},'' {\em Europhys.
		Lett.}, vol.~67, pp.~390--396, Aug. 2004.
	
	\bibitem{vanderPoel:JFM2013}
	E.~P. van~der Poel, R.~J. A.~M. Stevens, and D.~Lohse, ``{Comparison between
		two- and three-dimensional Rayleigh{\textendash}B{\'e}nard convection},''
	{\em J. Fluid Mech.}, vol.~736, pp.~177--194, Nov. 2013.
	
	\bibitem{Pandey:Pramana2016}
	A.~Pandey, M.~K. Verma, A.~G. Chatterjee, and B.~Dutta, ``{Similarities between
		2D and 3D convection for large Prandtl number},'' {\em Pramana-J. Phys.},
	vol.~87, p.~13, June 2016.
	
	\bibitem{Graham:JFM1975}
	E.~Graham, ``Numerical simulation of two-dimensional compressible convection,''
	{\em J. Fluid Mech.}, vol.~70, no.~4, pp.~689--703, 1975.
	
	\bibitem{Hurlburt:AJ1984}
	N.~E. Hurlburt, J.~Toomre, and J.~M. Massaguer, ``Two-dimensional compressible
	convection extending over multiple scale heights,'' {\em Astrophys. J.},
	vol.~282, pp.~557--573, 1984.
	
	\bibitem{Anderson:CFD1992}
	J.~D. Anderson, ``Governing equations of fluid dynamics,'' {\em Computational
		Fluid Dynamics: An Introduction}, pp.~15--51, 1992.
	
	\bibitem{cupy_learningsys2017}
	R.~Okuta, Y.~Unno, D.~Nishino, S.~Hido, and C.~Loomis, ``Cu{P}y: A
	{N}um{P}y-{C}ompatible {L}ibrary for {NVIDIA GPU C}alculations,'' in {\em
		Proceedings of Workshop on Machine Learning Systems (LearningSys) in The
		Thirty-first Annual Conference on Neural Information Processing Systems
		(NIPS)}, 2017.
	
	\bibitem{mpi4py}
	L.~Dalc{\'\i}n, R.~Paz, M.~Storti, and J.~D’El{\'\i}a, ``{MPI} for {P}ython:
	Performance improvements and {MPI}-2 extensions,'' {\em J. Parallel Distrib.
		Comput.}, vol.~68, no.~5, pp.~655--662, 2008.
	
	\bibitem{numpy}
	C.~R. Harris, K.~J. Millman, S.~J. Van Der~Walt, R.~Gommers, P.~Virtanen,
	D.~Cournapeau, E.~Wieser, J.~Taylor, S.~Berg, N.~J. Smith, {\em et~al.},
	``Array programming with {N}um{P}y,'' {\em Nature}, vol.~585, no.~7825,
	pp.~357--362, 2020.
	
	\bibitem{Stevens:JFM2010}
	R.~J. A.~M. Stevens, R.~Verzicco, and D.~Lohse, ``{Radial boundary layer
		structure and Nusselt number in Rayleigh{\textendash}B{\'e}nard
		convection},'' {\em J. Fluid Mech.}, vol.~643, pp.~495--507, Jan. 2010.
	
	\bibitem{Grotzbach:JCP1983}
	G.~Gr{\"o}tzbach, ``Spatial resolution requirements for direct numerical
	simulation of the {R}ayleigh-{B}{\'e}nard convection,'' {\em J. Comput.
		Phys.}, vol.~49, no.~2, pp.~241--264, 1983.
	
	\bibitem{Mishra:JFM2011}
	P.~K. Mishra, A.~K. De, M.~K. Verma, and V.~Eswaran, ``{Dynamics of
		reorientations and reversals of large-scale flow in
		Rayleigh{\textendash}B{\'e}nard convection},'' {\em J. Fluid Mech.},
	vol.~668, pp.~480--499, 2011.
	
	\bibitem{Landau:book:Fluid}
	L.~D. Landau and E.~M. Lifshitz, {\em {Fluid Mechanics}}.
	\newblock Course of Theoretical Physics, Oxford: Elsevier, 2nd~ed., 1987.
	
	\bibitem{movie}
	``\textit{The movies are available
		\href{https://www.youtube.com/playlist?list=PLzzEvSGvmTX9C74-sPEs_R6V3nEJelXee}{here}}.''
	YouTube, Aug 2024.
	
	\bibitem{Wu:PRA1991}
	X.-Z. Wu and A.~Libchaber, ``Non-boussinesq effects in free thermal
	convection,'' {\em Phys. Rev. A}, vol.~43, pp.~2833--2839, Mar 1991.
	
	\bibitem{Bhattacharya:PF2019}
	S.~Bhattacharya, R.~Samtaney, and M.~K. Verma, ``{Scaling and spatial
		intermittency of thermal dissipation in turbulent convection},'' {\em Phys.
		Fluids}, vol.~31, p.~075104, July 2019.
	
	\bibitem{Mishra:PRE2010}
	P.~K. Mishra and M.~K. Verma, ``{Energy spectra and fluxes for
		Rayleigh-B{\'e}nard convection},'' {\em Phys. Rev. E}, vol.~81, p.~056316,
	May 2010.
	
	\bibitem{Pandey:PRE2016}
	A.~Pandey, A.~Kumar, A.~G. Chatterjee, and M.~K. Verma, ``{Dynamics of
		large-scale quantities in Rayleigh-B{\'e}nard convection},'' {\em Phys. Rev.
		E}, vol.~94, p.~053106, Nov. 2016.
	
	\bibitem{vanderPoel:PRE2014}
	E.~P. Van Der~Poel, R.~Ostilla-M{\'o}nico, R.~Verzicco, and D.~Lohse, ``Effect
	of velocity boundary conditions on the heat transfer and flow topology in
	two-dimensional {R}ayleigh-{B}{\'e}nard convection,'' {\em Phys. Rev. E},
	vol.~90, no.~1, p.~013017, 2014.
	
	\bibitem{Verma:PRE2012}
	M.~K. Verma, P.~K. Mishra, A.~Pandey, and S.~Paul, ``{Scalings of field
		correlations and heat transport in turbulent convection},'' {\em Phys. Rev.
		E}, vol.~85, p.~016310, Jan. 2012.
	
\end{thebibliography}

\end{document}